\documentclass[12pt]{article}
\usepackage{fullpage}
\usepackage{epsf}
\usepackage{epsfig}
\usepackage{amsthm}
\usepackage{amsfonts}
\usepackage{color}
\usepackage{amsmath}
\usepackage{setspace}
\usepackage{natbib}
\usepackage{hyperref}
\usepackage{algorithmic}
\usepackage[boxed]{algorithm}
\usepackage[usenames, dvipsnames]{xcolor}

\begin{document}

\newcommand{\bm}[1]{\mbox{\boldmath $#1$}}
\newcommand{\mb}[1]{#1}%{\mathbf{#1}}
\newcommand{\bE}[0]{\mathbb{E}}
\newcommand{\bV}[0]{\mathbb{V}\mathrm{ar}}
\newcommand{\bP}[0]{\mathbb{P}}
\newcommand{\ve}[0]{\varepsilon}
\newcommand{\mN}[0]{\mathcal{N}}
\newcommand{\iidsim}[0]{\stackrel{\mathrm{iid}}{\sim}}
\newcommand{\NA}[0]{{\tt NA}}
\newcommand{\cB}{\mathcal{B}}
\newcommand{\R}{\mathbb{R}}
\newcommand{\Rp}{\R_+}

\title{\vspace{-1cm}  Bayesian optimization under mixed constraints 
with a slack-variable augmented Lagrangian}

\author{Victor Picheny\thanks{Institut National de Recherche Agronomique--Centre
de Toulouse, France} 
\and
Robert B.~Gramacy\thanks{Corresponding author: The University of
Chicago Booth School of Business, 5807 S.~Woodlawn Ave., Chicago IL, 60605;
\href{mailto:rbgramacy@chicagobooth.edu}{\tt rbgramacy@chicagobooth.edu}}
\and Stefan M.~Wild\thanks{Mathematics and Computer Science Division,
  Argonne National Laboratory, Lemont, IL 60439}
\and S\'ebastien Le~Digabel\thanks{{GERAD} and D\'epartement de
  Math\'ematiques et G\'enie Industriel, \'Ecole Polytechnique de
  Montr\'eal, Montr\'eal, QC H3C 3A7, Canada}}
  % Not needed for noncorresponding?
%   ; \href{https://www.gerad.ca/Sebastien.Le.Digabel}{\tt 
% www.gerad.ca/Sebastien.Le.Digabel}}}

\maketitle

\begin{abstract}
 An augmented Lagrangian (AL) can convert a constrained optimization problem
 into a sequence of simpler (e.g., unconstrained) problems, which are then
 usually solved with local solvers. Recently, surrogate-based Bayesian
 optimization (BO) sub-solvers have been successfully deployed in the AL
 framework for a more global search in the presence of inequality constraints;
 however, a drawback was that expected improvement (EI) evaluations relied on
 Monte Carlo. Here we introduce an alternative slack variable AL, and show
 that in this formulation the EI may be evaluated with library routines. The
 slack variables furthermore facilitate equality as well as inequality
 constraints, and mixtures thereof. We show how our new slack ``ALBO'' compares
 favorably to the original. Its superiority over conventional alternatives is
 reinforced on several mixed constraint examples.
\end{abstract}

\section{Introduction}
\label{sec:intro}

{\em Bayesian optimization} (BO), as applied to so-called {\em blackbox}
objectives, is a modernization of 1970-80s statistical response surface
methodology for sequential design~\citep{mockus:tiesis:zilinskas:1978,box:draper:1987,mockus:1989,mockus:1994}. In BO, nonparametric (Gaussian) processes~\citep[GPs,][]{rasmu:will:2006} provide flexible response surface fits.
Sequential design decisions, so-called {\em acquisitions}, judiciously balance
exploration and exploitation in search for global optima. For a review of GP
surrogate modeling and optimization in the context of computer experiments,
see~\citet{sant:will:notz:2003},~\citet{booker:etal:1999} and~\citet{bingham:etal:2014}.
For a machine learning perspective, see~\citet{brochu:cora:defreitas:2010}, and~\citet{boyle:2007}.
Until recently, most works in these literatures have
focused on unconstrained optimization.

Many interesting problems contain constraints, typically
specified as equalities or inequalities:
\begin{equation}
\min_x \left\{
f(x)
:
g(x) \leq 0, h(x) = 0, x \in \cB
\right\},
\label{eq:ineqprob}
\end{equation}
where $f:\R^d
\rightarrow \R$ is a scalar-valued objective function, and $g :
\R^d \rightarrow \R^m$ and $h : \R^d \rightarrow \R^p$  are  
vector-valued constraint functions taken 
componentwise (i.e., $g_j(x) \leq 0$, $j=1,\dots,m$; 
% $h_k(x) = 0$, $k=1,\dots,p$ and all $m+p$
$h_k(x) = 0$, and $k=1,\dots,p$).
We differentiate these constraints from $\cB \subset \R^d$, which is a 
known (not blackbox) set, typically containing bound constraints; here we 
assume that $\cB$ is a bounded hyperrectangle.\footnote{The presumption 
throughout is that a 
solution of~\eqref{eq:ineqprob} exists so that the feasible region $\{x \in
\R^d: g(x)\leq 0, \, h(x)= 0\} \cap \cB$ is nonempty; however,
the algorithms we describe can provide reasonable approximate
solutions when the feasible region is empty, 
so long as the modeling assumptions 
(e.g., smoothness and stationarity)
on $f$, $g$, and $h$, and constraint qualifications (see, e.g.,~\citep{nocedal2006no}) are not egregiously violated.}

The typical setup treats $f$, $g$, and $h$ as a ``joint'' blackbox, meaning
that providing $x$ to a single computer code reveals $f(x)$, $g(x)$, and
$h(x)$ simultaneously, often at great computational expense. A common special
case treats $f(x)$ as known (e.g., linear); however, the problem is still hard
when $g(x)\leq 0$ defines a nonconvex valid region. A tacit goal in this setup
is to minimize the number of blackbox runs required to solve the problem, and
a common meter of progress tracks the best valid value of the objective (up to
a relaxing threshold on the equality constraints $h$) at increasing
computational budgets (\# of evaluations, $n$).

Not many algorithms target global solutions to this general, constrained
blackbox optimization problem.  Statistical methods are acutely few.  We know
of no methods from the BO literature natively accommodating equality
constraints, let alone mixed (equality and inequality) ones.
\citet{scho:welc:jone:1998} describe how their expected improvement (EI)
heuristic can be extended to multiple inequality constraints by multiplying by
an estimated probability of constraint satisfaction. Here, we call this {\em
expected feasible improvement (EFI)}. EFI has recently been revisited by
several authors~\citep{snoek:etal:2012,gelbart:etal:2014,gardner:etal:2014}.
However, the technique has pathological behavior in otherwise idealized
setups~\citep{rejoinder}, which is related to a so-called
``decoupled'' pathology~\citep{gelbart:etal:2014}.  Some recent
information-theoretic alternatives have shown promise in the inequality
constrained setting~\citep{hernandez:etal:2015,picheny:2014}; tailored
approaches have shown promise in more specialized constrained
optimized setups~\citep{gramacy:lee:2011,will:sant:notz:lehm:2010,parr:etal:2012}.

We remark that any problem with equality constraints can be ``transformed'' to
inequality constraints only, by applying $h(x)\leq 0$ and $h(x) \geq 0$
simultaneously. However, the effect of such a reformulation is rather
uncertain.  It puts double-weight on equalities and violates certain
regularity (i.e., constraint qualification~\citep{nocedal2006no}) conditions.
Numerical issues have been reported in empirical
evaluations~\citep{sasena:2002}. In our own empirical work
[Section~\ref{sec:empirical}] we find unfavorable performance.

In this paper we show how a recent technique~\citep{GrGrLedLeeRaWeWi2016} for
BO under inequality constraints is naturally enhanced to
handle equality constraints, and therefore mixed ones too. The method involves
converting inequality constrained problems into a sequence of simpler
subproblems via the augmented Lagrangian~\citep[AL,][]{Bertsekas82}).  AL-based solvers can, under certain regularity
conditions, be shown to converge to locally optimal solutions that satisfy the
constraints, so long as the sub-solver converges to local
solutions. By deploying modern BO on the subproblems, as opposed to the usual
local solvers, the resulting meta-optimizer is able to find better, less local
 solutions with fewer evaluations of
the expensive blackbox, compared to several classical and statistical
alternatives.  Here we dub that method ALBO.

To extend ALBO to equality constraints, we suggest the opposite
transformation to the one described above: we convert inequality
constraints into equalities by introducing {\em slack variables}. In the
context of earlier work with the AL, via conventional solvers, this is rather
textbook~\citep[][Ch.~17]{nocedal2006no}. Handling the
inequalities in this way leads naturally to solutions for mixed constraints
and, more importantly, dramatically improves the original {\em
inequality-only} version.  In the original (non-slack) ALBO setup, the density
and distribution of an important composite random predictive quantity is not
known in closed form. Except in a few particular cases~\citep{discussion4},
calculating EI and related quantities under the AL required Monte Carlo
integration, which means that acquisition function evaluations are
computationally expensive, noisy, or both. A reformulated slack-AL version
emits a composite that has a known distribution, a so-called weighted
non-central Chi-square (WNCS) distribution.  We show that, in that setting, EI
calculations involve a simple 1-d integral via ordinary quadrature. Adding
slack variables increases the input dimension of the optimization subproblems,
but only artificially so.  The effects of expansion can be mitigated through
optimal default settings, which we provide.

The remainder of the paper is organized as follows. Section~\ref{sec:review}
outlines the components germane to the ALBO approach: AL, Bayesian surrogate
modeling, and acquisition via EI.  Section~\ref{sec:slack} contains the bulk
of our methodological contribution, reformulating the AL with slack variables
and showing how the EI metric may be calculated in closed form, along with
optimal default slack settings, and open-source software. Implementation
details are provided in our Appendix. Section~\ref{sec:empirical} provides
empirical comparisons, and Section~\ref{sec:discuss} concludes.

\section{A review of relevant concepts: EI and AL}
\label{sec:review}

\subsection{Expected improvement}
\label{sec:ei}

The canonical acquisition function in BO is {\em expected
improvement}~\citep[EI,][]{jones:schonlau:welch:1998}. Consider a surrogate $f^n(x)$, trained on
$n$ pairs $(x_i, y_i = f(x_i))$ emitting Gaussian predictive equations with
mean $\mu^n(x)$ and standard deviation $\sigma^{n}(x)$. Define $f^n_{\min} =
\min_{i=1,\ldots,n} y_i$, the smallest $y$-value seen so
far, and let $I(\mb{x}) = \max\{ 0, f^n_{\min} - Y(x) \}$
be the % (potential) 
{\em improvement} at $x$.  $I(x)$ is largest when $Y(\mb{x}) \sim f^n(x)$ has
substantial distribution below $f^n_{\min}$. 
The expectation of $I(x)$ over $Y(x)$ has a convenient closed form,
revealing balance between exploitation ($\mu^n(x)$ under $f^n_{\min}$) and
exploration (large $\sigma^{n}(x)$):
\begin{equation}
\bE\{I(x)\} = (f^n_{\min} - \mu^n(x)) \Phi\left(
\frac{f^n_{\min} - \mu^n(x)}{\sigma^n(x)}\right)
+ \sigma^n(x) \phi\left(
\frac{f^n_{\min} - \mu^n(x)}{\sigma^n(x)}\right),
\label{eq:ei}
\end{equation}
where $\Phi$ ($\phi$) is the standard normal cdf (pdf). Accurate, 
approximately Gaussian predictive equations are provided by many statistical
models~\citep[GPs,][]{rasmu:will:2006}. 

When the predictive equations are not Gaussian, Monte Carlo schemes---sampling
$Y(x)$ variates from $f^n(x)$ and averaging the resulting $I(x)$
values---offer a suitable, if computationally intensive, alternative.
Yet most of the theory applies only to the Gaussian case. Under certain
regularity conditions~\citep[e.g.,][]{bull:2011}, one can show that sequential designs derived by
repeating the process, generating $(n+1)$-sized designs from $n$-sized ones,
contain a point $x_i$ whose associated $y_i=f(x_i)$ is arbitrarily close to 
the true minimum
objective value, that is, the algorithm converges.

\subsection{Augmented Lagrangian framework}
\label{sec:auglag}

 Although several authors have suggested extensions to EI for
constraints, the BO literature has primarily focused on unconstrained
problems. The range of constrained BO options was recently extended by
borrowing an apparatus from the mathematical optimization literature,
the {\em augmented Lagrangian}, allowing unconstrained methods to be
adapted to constrained problems. The AL, as a device for solving problems with
inequality constraints (no $h(x)$ in Eq.~\eqref{eq:ineqprob}), may be defined as
\begin{equation}
L_A(\mb{x};\lambda, \rho) = f(\mb{x}) +\lambda^\top g(\mb{x}) + \frac{1}{2\rho}
\sum_{j=1}^m \max \left\{0,g_j(x) \right\}^2,
\label{eq:la}
\end{equation}
where $\rho>0$ is a \emph{penalty parameter} on constraint violation and
$\lambda\in\Rp^m$ serves as a \emph{Lagrange multiplier}. AL methods 
are
iterative, involving a particular sequence of $(\mb{x};\lambda, \rho)$.
Given the current values 
$\rho^{k-1}$ and $\lambda^{k-1}$,
one approximately solves the subproblem
\begin{equation}
\min_x \left\{ L_A(x;\lambda^{k-1},\rho^{k-1}) : x\in \cB\right\},
 \label{eq:auglagsp}
 \end{equation}
via a conventional (bound-constrained) solver.  The parameters $(\lambda, 
\rho)$ are updated depending
on the nature of the solution found, and the process repeats. The particulars
in our setup are provided in Alg.~1; for more details see~\citep[][Ch.~17]{nocedal2006no}. Local convergence % (to valid solutions) 
is guaranteed under relatively mild conditions involving the choice of
subroutine solving~\eqref{eq:auglagsp}.  Loosely, all that is required is that
the solver ``makes progress'' on the subproblem.  In contexts where
termination depends more upon computational budget than on a measure of
convergence, as in many BO problems, that added flexibility is welcome.

\begin{algorithm}[ht!]
\begin{algorithmic}[1]
\REQUIRE $\lambda^0\geq 0$, $\rho^0>0$
\FOR{$k=1, 2, \ldots $}
\STATE Let $x^k$ (approximately) solve (\ref{eq:auglagsp})
\STATE Set $\lambda_j^k =
\max\left(0,\lambda_j^{k-1}+\frac{1}{\rho^{k-1}} g_j(x^k) \right)$, $j=1,
\ldots, m$
\STATE If $g(x^k)\leq 0$, set $\rho^k=\rho^{k-1}$; otherwise, set
$\rho^k=\frac{1}{2}\rho^{k-1}$
\ENDFOR
\end{algorithmic}
\caption{Basic augmented Lagrangian method.}
\label{alg:baseal}
\end{algorithm}

However, the AL does not typically enjoy global scope.  The local minima found
by the method are sensitive to initialization---of starting choices for
$(\lambda^0, \rho^0)$ or $x^0$; local searches in iteration $k$ are usually 
started from $x^{k-1}$. However, this dependence is broken 
when statistical surrogates drive search for solutions to the subproblems.
%\citet{GrGrLedLeeRaWeWi2016} proposed separately modeling the components
% of~eqref{eq:auglagsp}.\footnote{That is, as opposed to modeling the whole thing with one
% single surrogate.  They describe how that setup suffers from pathologies such
% as kinks (due to the $\max$) and amplifications (due to the square) that
% challenge the typical GP surrogate modeling framework.}  
Independently fit GP surrogates, $f^n(x)$ for the objective and $g^n(x) =
(g_1^n(x), \dots, g_m^n(x))$ for the constraints, yield predictive
distributions for $Y_f^n(\mb{x})$ and $Y_g^n(x) = (Y_{g_1}^n(\mb{x}), \dots,
Y_{g_m}^n(\mb{x}))$. Dropping the $n$ superscripts, the AL composite random
variable
\begin{equation}\label{eq:Yorig}
 Y(x) = Y_f(x) + \lambda^\top Y_g(x) + \frac{1}{2\rho} \sum_{j=1}^m \max\{0, 
Y_{g_j}(x)\}^2
\end{equation}
 can serve as a surrogate for~\eqref{eq:la}; however, it is difficult to
deduce the full distribution of this variable from the components of $Y_f$ and
$Y_g$, even when those are independently Gaussian. While its mean is available in
closed form, EI requires Monte Carlo.

\section{A novel formulation involving slack variables}
\label{sec:slack}

An equivalent formulation of~\eqref{eq:ineqprob} involves introducing slack variables, $s_j$, for
$j=1,\dots,m$ (i.e., one for each inequality constraint $g_j(x)$), and 
converting
the mixed constraint problem~\eqref{eq:ineqprob} into one with 
only equality constraints (plus bound constraints for $s_j$):
\begin{equation}
 g_j(x) - s_j = 0, \quad s_j \in \Rp, \quad \text{for } j=1,\dots, m.
\label{eq:slack}
\end{equation}
The input dimension of the problem is increased from $d$ to
$d+m$ with the introduction of the $m$ (positive) slack ``inputs'' $s$. 
% In practice, the
% inequality constraints on $s_j$ are subsumed into $\mathcal B \subseteq
% \R^{d+m}$. They are unbounded rather than convex in the latter $m$ dimensions,
% which means that some care is required---further discussion is delayed to
% Section~\ref{sec:sopt}, after discussing our (AL-based) strategy.

Reducing a mixed constraint problem to one involving only equality 
and bound constraints is
valuable insofar as one has good solvers for those problems.  Indeed, the AL
method is an attractive option here, but some adaptation is required. Suppose,
for the moment, that the original problem~\eqref{eq:ineqprob} has no equality
constraints (i.e., $p=0$).  In this case, a slack variable-based AL method is
readily available---as an alternative to the one described in Section~\ref{sec:auglag}.
Although we frame it as an ``alternative'', some in the 
mathematical optimization community
would describe this as the standard version~\citep[see, e.g.,][Ch.~17]{nocedal2006no}. The AL for this setup is given as follows:
\begin{equation}
L_A(\mb{x}, s;\lambda_g,\rho) = f(\mb{x}) + \lambda^\top \left(g(\mb{x})\!+\!s \right)
+ \frac{1}{2\rho} \sum_{j=1}^{m} \left(g_j(x)\!+\!s_j\right)^2.
\label{eq:aleq}
\end{equation}
This formulation is more convenient than~\eqref{eq:la} because the ``max'' is
missing, but the extra slack variables mean solving a higher ($d+m$)
dimensional subproblem compared to~\eqref{eq:auglagsp}.  In
Section~\ref{sec:alg} we address this issue, and further detail adjustments to
Alg.~\ref{alg:baseal} for the slack variable case.%Some changes to the
%updating scheme in Alg.~1 are required, which we outline in Section~\ref{sec:alg}.

% Observe that the role of the slack variable, whether in the problem
% specification~\eqref{eq:slack} or the AL~\eqref{eq:aleq}, is to allow
% solutions to be found off of the constraint boundaries $\{x\in \R^m : g_j(x) = 
% 0\}$.  In cases where $\hat{s}_j = 0$, the
% inequality constraint $j$ is said to be {\em binding}.  Therefore, by
% inverting the reasoning behind the reformulation of Eq.~\eqref{eq:slack}, one
% can equivalently represent an equality constraint as an inequality constraint
% where binding is enforced through a fixed setting of the slack variable, $s_j
% = 0$.  This means that the 
The AL can be expanded to handle equality (and thereby mixed constraints) as
follows:
\begin{equation}
L_A(\mb{x}, s;\lambda_g,\lambda_h, \rho) = f(\mb{x}) + \lambda_g^\top \left(g(\mb{x})\!+\!s \right)+  \lambda_h^\top h(\mb{x}) + \frac{1}{2\rho}\!
\left[ \sum_{j=1}^m \left(g_j(x)\!+\!s_j\right)^2 + \sum_{k=1}^p  h_k(x)^2\right]\!.
\label{eq:straw}
\end{equation}
% with tacit $s_k = 0$ in the rightmost sum above. 
Defining $c(x) := \left[g(x)^\top, h(x)^\top\right]^\top\!\!$, $\lambda :=
\left[\lambda_g^\top, \lambda_h^\top \right]^\top\!\!$, and enlarging the dimension of $s$ 
with the understanding that $s_{m+1}=\cdots=s_{m+p}=0$, leads to a streamlined AL
for mixed constraints
\begin{equation}
L_A(\mb{x}, s; \lambda, \rho) = f(\mb{x}) + \lambda^\top \left(c(\mb{x}) + s \right)
+ \frac{1}{2\rho} \sum_{j=1}^{m+p} \left(c_j(x) + s_j\right)^2,
\label{eq:ela2}
\end{equation}
with $\lambda\in \R^{m+p}$.
% allowing an equality-constraint AL to be deployed in a mixed constraint setting. 
A non-slack AL formulation~\eqref{eq:la} may be written analogously as
\[
L_A(\mb{x};\lambda_g, \lambda_h, \rho) = f(\mb{x}) + \lambda_g^\top g(\mb{x}) +  \lambda_h^\top h(\mb{x}) + \frac{1}{2\rho}
\left[ \sum_{j=1}^m \max \left\{0,g_j(x) \right\}^2 + \sum_{k=1}^p  
h_k(x)^2\right],
\]
 with $\lambda_g\in\Rp^m$ and $\lambda_h\in\R^p$. Eq.~\eqref{eq:ela2}, by
contrast, is easier to work with because it is a smooth quadratic in the
objective ($f$) and constraints ($c$).
% \vic{I haven't found if the $\lambda_h$ must be positive or not. We need to check that, as well
% as in our implementation.}   
In what follows, we show that~\eqref{eq:ela2} facilitates calculation of
important quantities like EI, in the GP-based BO framework, via a library
routine. So slack variables not only facilitate mixed
constraints in a unified framework, they also lead
to a more efficient handling of the original inequality (only) constrained
problem.

\subsection{Distribution of the slack-AL composite}
If $Y_f$ and $Y_{c_1}, \dots, Y_{c_{m+p}}$ represent random predictive variables
from ${m+p}+1$ surrogates fitted to $n$ realized objective and constraint
evaluations, then the analogous slack-AL random variable is
\begin{equation}
 Y(x, s) = Y_f(x) + \sum_{j=1}^{m+p} \lambda_j (Y_{c_j}(x) + s_j) + 
 \frac{1}{2\rho} \sum_{j=1}^{m+p} (Y_{c_j}(x) + s_j)^2.
\label{eq:sal}
\end{equation} 
% Above, we write out the dot-product $\lambda^\top(Y_c + s)$ as a sum to ease
% future calculations. 
As in the original AL formulation, the mean of this RV has a simple closed
form in terms of the means and variances of the surrogates, regardless of the
form of predictive distribution. In the conditionally Gaussian case, we can
derive the full distribution of the slack-AL variate~\eqref{eq:sal} in closed
form.  % can be recognized as a quadratic form of Gaussian RVs, which as we now show,
% has a generalized chi-square distribution.
%
Toward that aim, we re-develop the composite $Y$ as follows:
\begin{align}
 Y(x, s) &= Y_f(x) + \sum_{j=1}^{m+p} \lambda_j s_j + 
  \frac{1}{2\rho} \sum_{j=1}^{m+p} s_j^2 + \frac{1}{2\rho} 
  \sum_{j=1}^{m+p} \left[ 2 \lambda_j \rho Y_{c_j}(x) + 2 s_j Y_{c_j}(x) 
+ Y_{c_j}(x)^2 \right] 
  \nonumber \\
     &= Y_f(x) + \sum_{j=1}^{m+p} \lambda_j s_j + \frac{1}{2\rho} 
\sum_{j=1}^{m+p} s_j^2 + \frac{1}{2\rho} \sum_{j=1}^{m+p} \left[\left( \alpha_j 
+ Y_{c_j}(x) \right)^2 - \alpha_j^2 \right],
     \nonumber
\intertext{with $\alpha_j=  \lambda_j\rho + s_j$. Now decompose the $Y(x,s)$ into 
a sum of three quantities:}
Y(x, s) &= Y_f(x) + r(s) + \frac{1}{2\rho} W(x, s), \quad \mbox{with} \label{eq:Y} \\
r(s) &= \sum_{j=1}^{m+p} \lambda_j s_j + \frac{1}{2\rho} 
\sum_{j=1}^{m+p} s_j^2 - \frac{1}{2\rho} \sum_{j=1}^{m+p} \alpha_j^2 
\quad \mbox{and} \quad
W(x, s) = \sum_{j=1}^{m+p} \left( \alpha_j + Y_{c_j}(x) \right)^2. \nonumber
\end{align}
Using $Y_{c_j} \sim \mN \left(\mu_{c_j}(x), \sigma^2_{c_j}(x) \right)$, i.e.,
leveraging Gaussianity, $W$ can be written as
\begin{equation}\label{eq:Z}
 W(x, s) =  \sum_{j=1}^{m+p} \sigma^2_{c_j}(x)  X_j(x,s), \quad \mbox{with } 
    X_j(x,s) \sim \chi^2 \left(\mathrm{dof}\!=\!1, \delta\!=\! 
    \left( \frac{\mu_{c_j}(x) + \alpha_j}{\sigma_{c_j}(x)}\right)^2 \right). 
\hfill
\end{equation}
% 
% 
% \begin{align}
%    W(x, s) &= \sum_{j=1}^m Z_j^2, && \mbox{with } &
%    Z_j &\sim \mN\left(\mu_{c_j}(x) + \alpha_j, \sigma^2_{c_j}(x) \right) \label{eq:Z} \\
%    &=  \sum_{j=1}^m \sigma^2_{c_j}(x) \bar{Z}_j^2, && \mbox{with } & 
%    \bar{Z}_j &\sim \mN\left(\frac{\mu_{c_j}(x) + \alpha_j}{\sigma_{c_j}(x)}, 1 \right) \nonumber \\
%     &=  \sum_{j=1}^m \sigma^2_{c_j}(x)  X_j, && \mbox{with } &
%     X_j &\sim \chi^2 \left(\mathrm{dof}=1, \delta= 
%     \left( \frac{\mu_{c_j}(x) + \alpha_j}{\sigma_{c_j}(x)}\right)^2 \right).
%     \nonumber
% \end{align}
The line above is the expression of a weighted sum of non-central chi-square
(WSNC) variates. Each of the ${m+p}$ variates involves a unit
degrees-of-freedom (dof) parameter, and a non-centrality parameter $\delta$. A
number of efficient methods exist for evaluating the density, distribution,
and quantile functions of WSNC random variables~\citep[e.g.,][]{davies:1980,duchense:lafaye:2010}. The {\sf R} packages {\tt
CompQuadForm}~\citep{CompQuadForm-Manual} and {\tt sadists}~\citep{sadists-Manual} provide library implementations. Details and code are
provided in Appendix~\ref{sec:wsnc}.

Some constrained optimization problems involve an {\em a priori} known
objective $f(x)$. In that case, referring back to~\eqref{eq:Y}, we are done:
the distribution of $Y(x,s)$ is WSNC (as in~\eqref{eq:Z}) shifted by a known
quantity $f(x) + r(s)$, determined by particular choices of inputs $x$ and
slacks $s$.  When $Y_f(x)$ is conditionally Gaussian we have that
$\tilde{W}(x,s) = Y_f(x) + \frac{1}{2\rho} W(x,s)$ is the weighted sum of a
Gaussian and WNCS variates, a problem that is again well-studied. The same
methods cited above are easily augmented to handle that case---see Appendix~\ref{sec:wsnc}.

\subsection{Slack-AL expected improvement}
\label{sec:sei}

Evaluating EI at candidate $(x,s)$ locations under the
AL-composite involves working with $\mathrm{EI}(x, s) = \mathbb{E} \left[
\left( y_{\min}^n - Y(x, s) \right)
    \mathbb{I}_{\{Y(x, s) \leq y_{\min}^n\}}  \right]$,
given the current minimum $y^n_{\min}$ of the AL over all $n$ runs. When  $f(x)$
is known, let $w^n_{\min}(x,s) =  2\rho \left(y^n_{\min} - f(x) - r(s)
\right)$ absorb all of the non-random quantities involved in the EI
calculation. Then, with $D_W(\cdot;x,s)$ denoting the distribution of
$W(x,s)$,
\begin{eqnarray}
 \mathrm{EI}(x, s) &=& \frac{1}{2\rho} \mathbb{E} 
  \left[ \left( w^n_{\min}(x,s) - W(x,s) \right) \mathbb{I}_{W(x,s) \leq 
w_{\min}(x,s)}  \right] \nonumber\\
&=& \frac{1}{2\rho} \int_{-\infty}^{w^n_{\min}(x,s)} D_W(t;x,s) dt
= \frac{1}{2\rho} \int_{0}^{w^n_{\min}(x,s)} D_W(t;x,s) dt\label{eq:EIs}
\end{eqnarray}
if $w^n_{\min}(x,s) \geq 0$ and zero otherwise.
That is, the EI boils down to integrating the distribution function of
$W(x,s)$ between 0 (since $W$ is positive) and $w^n_{\min}(x,s)$.  This is a
one-dimensional definite integral that is easy to approximate via quadrature;
details are in Appendix~\ref{sec:wsnc}.
Since $W(x,s)$ is quadratic in the
$Y_c(x)$ values, it is often the case, especially for smaller $\rho$-values in
later AL iterations, that $D_W(t;x,s)$ is zero over most of $[0,
w^n_{\min}(x,s)]$, simplifying the numerical integration. However, this has
deleterious impacts on search over $(x,s)$. 
%Note that exactly zero EI when $w^n_{\min}(x,s) \leq 0$ can have
%deleterious impacts on search over $(x,s)$. 
We discuss a useful adaptation for that case in Appendix~\ref{sec:open}.

When $f(x)$ is unknown and $Y_f(x)$ is conditionally normal, let 
$\tilde{w}^n_{\min}(s) = 2\rho \left(y^n_{\min} - r(s) \right)$.
Then,
\[
\mathrm{EI}(x, s) = \frac{1}{2\rho} \mathbb{E} 
  \left[ \left( \tilde{w}^n_{\min}(s) - \tilde{W}(x,s) \right) 
  \mathbb{I}_{\tilde{W}(x,s) \leq \tilde{w}^n_{\min}(s)}  \right] 
= \frac{1}{2\rho} \int_{-\infty}^{\tilde{w}^n_{\min}(s)} D_{\tilde{W}}(t;x,s) 
dt.
\]
Here the lower bound of the definite integral cannot be zero since
$Y_f(x)$ may be negative, and thus $\tilde{W}(x,s)$ may have non-zero
distribution for negative $t$-values.  This presents some challenges when it
comes to numerical integration via quadrature, although many library functions
allow indefinite bounds. We obtain better performance by supplying a
conservative finite lower bound, for example three standard deviations in
$Y_f(x)$, in units of the penalty ($2\rho$), below zero: $6 \rho \sigma_f(x)$. % \vic{not very clear} 
Detailed example implementations are provided Appendix~\ref{sec:wsnc}.

% \section{The new slack-AL algorithm}
\subsection{AL updates, optimal slacks, and other implementation notes}
\label{sec:alg}

The new slack-AL method is completed by describing when the
 subproblem~\eqref{eq:ela2} is deemed to be ``solved'' (step 2 in
 Alg.~\ref{alg:baseal}), how $\lambda$ and $\rho$ updated (steps 3--4). We
 terminate the BO search sub-solver after a single iteration as this matches
 with the spirit of EI-based search, whose choice of next location can be
 shown to be optimal, in a certain sense, if it is the final point being
 selected~\citep{bull:2011}. It also meshes well with an updating scheme
 analogous to that in steps 3--4: updating only when {\em no} actual
 improvement (in terms of constraint violation) is realized by that choice.
 For those updates, the analog for the mixed constraint setup is

\begin{quote}
step 2: Let $(x^k,s^k)$ approx.~solve 
$\min_{x,s}\!\left\{ 
L_A(x,s;\lambda^{k-1},\rho^{k-1}) : (x,s_{1:m})\in \tilde{\cB} \right\}$ \\ 
%$, s_{m+1:m+p} =  0\right\}$ 
step 3: $\lambda_j^k = \lambda_j^{k-1} + \frac{1}{\rho^{k-1}} (c_j(x^k) + 
s_j^k)$, for $j=1,
\dots, m+p$ \\ 
step 4: If $c_{1:m}(x^k)\leq 0$ and $|c_{m+1:m+p}(x^k)| \leq
\epsilon$, set $\rho^k\!\!=\!\!\rho^{k-1}$; else $\rho^k\!=\!\frac{1}{2}
\rho^{k-1}$
\end{quote}
\noindent Above, step 3 is the same as in Alg.~1 except
without the ``max'', and  with slacks augmenting the constraint
values.  The ``if'' statement in step 4 checks for validity at $x^k$,
deploying a threshold $\epsilon \geq 0$ on equality constraints;  further 
discussion of the threshold $\epsilon$ is deferred 
to Section~\ref{sec:empirical}, where we
discuss progress metrics under mixed constraints.  If validity holds at 
$(x^k,s^k)$, the current AL iteration
is deemed to have ``made progress'' and the penalty remains unchanged; otherwise
it is doubled.  An alternate formulation may check $|c_{1:m}(x^k)+s_{1:m}^k|
\leq \epsilon$.  We find that the version in step 4, above, is cleaner because 
it limits sensitivity to the choice of threshold $\epsilon$.  
In Appendix~\ref{sec:alparams} recommend initial $(\lambda^0, \rho^0)$ values 
which are analogous to the original, non-slack AL settings.

% These are trivial changes.  
{\bf Optimal choice of slacks:} The biggest difference between the original AL~\eqref{eq:la}
and slack-AL~\eqref{eq:ela2} is that the latter requires
searching over both $x$ and $s$, whereas the former involves only $x$-values.
In what follows we show that there are automatic choices for the $s$-values as
a function of the corresponding $x$'s, keeping the search space
$d$-dimensional, rather than $d+m$.

%\subsection{Optimal choice of slack variables}
\label{sec:sopt}

 For an observed $c_j(x)$ value,
%For any $x$-value, 
associated slack variables minimizing the AL~\eqref{eq:ela2} can be obtained analytically. 
Using the form of~\eqref{eq:Y},
observe that $\min_{s
\in \Rp^m} y(x,s)$ 
is equivalent to
$\min_{s\in \Rp^m} \sum_{j=1}^m 2\lambda_j \rho s_j + s_j^2 + 2 s_j c_j(x)$.
For fixed $x$, this is strictly convex in $s$. 
 Therefore, its unconstrained minimum can only
be its stationary point, which satisfies $0 = 2\lambda_j\rho + 2s_j^*(x) +
2c_j(x)$, for $j=1,\dots,m$. Accounting for the nonnegativity constraint, we
obtain the following optimal slack as a function of $x$:
\begin{equation}
s_j^*(x) = \max \left\{ 0, - \lambda_j\rho - c_j(x) \right\}, \qquad 
j=1,\dots,m. \label{eq:sopt}
\end{equation}
Above we write $s^*$ as a function of $x$ to convey that $x$
remains a ``free'' quantity in $y(x, s^*(x))$. Recall that slacks on
equality constraints are zero, $s_k(x) = 0$, $k=m+1,\dots,m+p$, for all
$x$.

In the blackbox $c(x)$ setting, $y(x, s^*(x))$ is only directly accessible at
the data locations $x_i$. At other $x$-values, however, the surrogates provide
a useful approximation. When $Y_c(x)$ is (approximately) Gaussian it is
straightforward to show that the optimal setting of the slack variables,
solving $\min_{s \in \Rp^m}
\mathbb{E} [ Y(x,s) ]$, are $s_j^*(x) = \max\{0, -\lambda_j \rho -
\mu_{c_j}(x)\}$, i.e., the same as~\eqref{eq:sopt} with a prediction 
$\mu_{c_j}(x)$ for
$Y_{c_j}(x)$ unknown $c_j(x)$ value.  Again, slacks on
the equality constraints are set to zero.

Other criteria can be used to choose slack variables. Instead of minimizing
the mean of the composite, one could maximize the EI. In Appendix~\ref{sec:optslack} we explain how this is of dubious practical value.  Compared to the settings
described above, searching over EI is both
more computationally intensive and provides near identical results in
practice.

%\subsection{Implementation notes}
\label{sec:optei}

{\bf Implementation notes:} Code supporting all methods in this manuscript is
provided in two open-source {\sf R} packages: {\tt laGP}~\citep{laGP} and {\tt
DiceOptim}~\citep{DiceOptim}, both on CRAN~\citep{cran:R}.  Implementation
details vary somewhat across those packages, due primarily to particulars of
their surrogate modeling capability and how they search the EI surface.  For
example, {\tt laGP} can accommodate a smaller initial design size because it
learns fewer parameters (i.e., has fewer degrees of freedom). {\tt DiceOptim}
uses a multi-start search procedure for EI, whereas {\tt laGP} deploys a
random candidate grid, which may optionally be ``finished'' with an L-BFGS-B
search.  Nevertheless, their qualitative behavior exhibits strong similarity.
Both packages also implement the original AL scheme (i.e., without slack
variables) updated~\eqref{eq:straw} for mixed constraints.  Further details
are provided in Appendix~\ref{sec:open}.

\section{Empirical comparison}
\label{sec:empirical}

Here we describe three test problems, each mixing challenging
elements from traditional unconstrained blackbox optimization benchmarks,
but in a constrained optimization format. We run our optimizers on these
problems 100 times under random initializations.  In the case of our GP
surrogate comparators, this initialization involves choosing random
space-filling designs. Our primary means of comparison is an averaged
(over the 100 runs) measure of progress defined by the best valid value of the
objective for increasing budgets (number of evaluations of the blackbox), $n$.

In the presence of equality constraints it is necessary to relax this
definition somewhat, as the valid set may be of measure zero.  In such cases
we choose a tolerance $\epsilon\geq 0$ and declare a solution to be ``valid''
when inequality constraints are all valid, and when $|h_k(x)| \leq \epsilon$ for
%! SW: Note I chanced ``< \epsilon'' to ``\leq epsilon'' above
all $k=1, \dots, p$.  In our figures we choose $\epsilon = 10^{-2}$; however,
the results are similar under stronger thresholds, with a higher 
variability over initializations.
% degree of Monte Carlo (MC) error.  
As finding a valid solution is, in itself, sometimes a difficult task, we
additionally report the proportion of runs that find valid and optimal
solutions as a function of budget, $n$, for problems with equality (and mixed)
constraints.
% 
% In several cases we provide alternate views into
% relative performance to better understand variability among the 100 MC
% re-initializations.  For example, we provide the proportion of runs that find
% valid and optimal solutions as a function of budget, $n$.

\subsection{An inequality constrained problem}
\label{sec:ineq}

We first revisit the ``toy'' problem from~\citet{GrGrLedLeeRaWeWi2016}, 
having a 2-d input space limited to the unit cube, 
a (known) linear objective, 
% and two inequality constraints 
with sinusoidal and quadratic inequality constraints (henceforth LSQ problem;
see the Appendix~\ref{sec:testpbs} for details). Figure~\ref{f:toy}
shows progress over repeated solves with a maximum budget of 40
 blackbox evaluations.
\begin{figure}
\centering
%\vspace{-0.4cm}
\includegraphics[scale=0.625,trim=0 20 0 40]{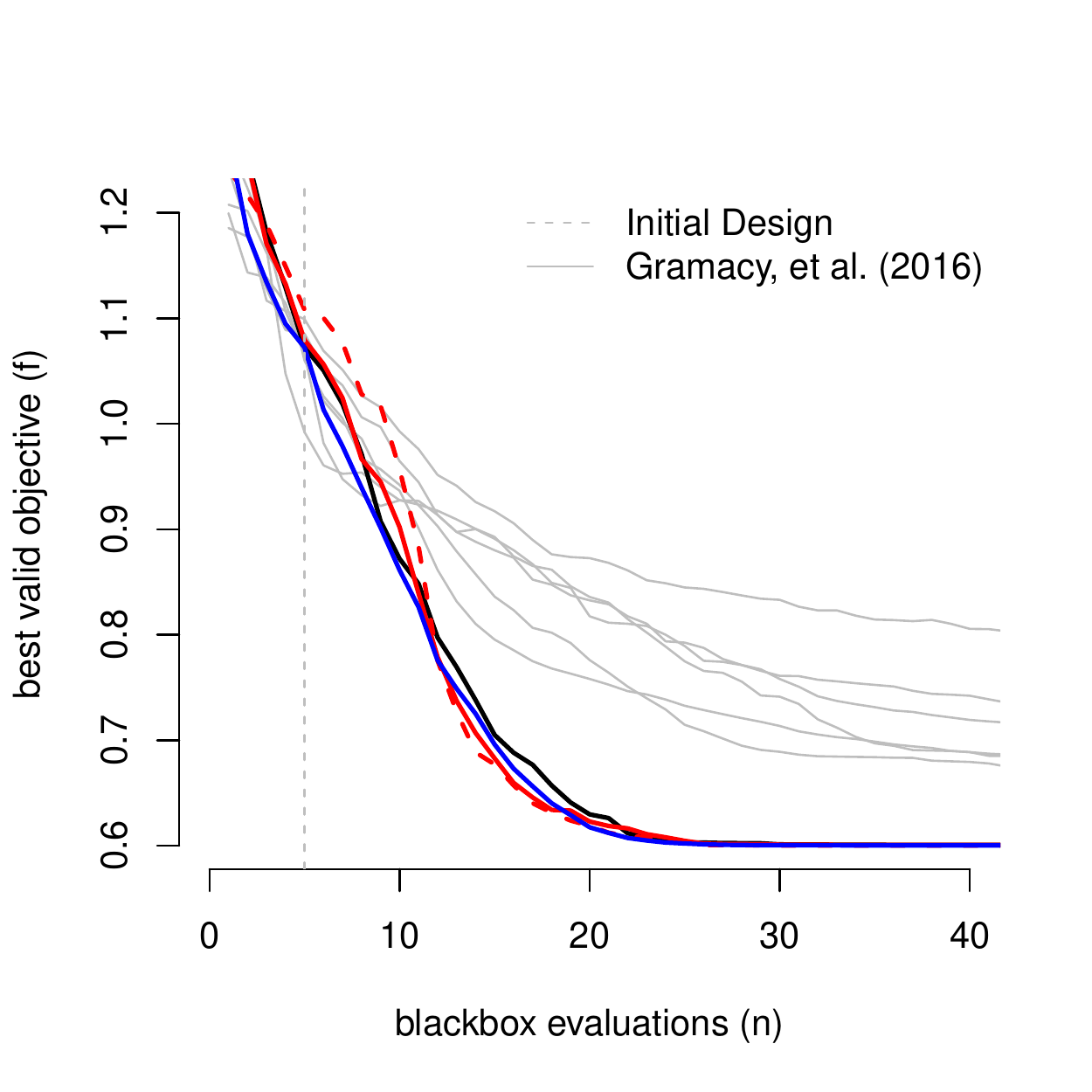} \hfill
\includegraphics[scale=0.625,trim=0 20 0 40]{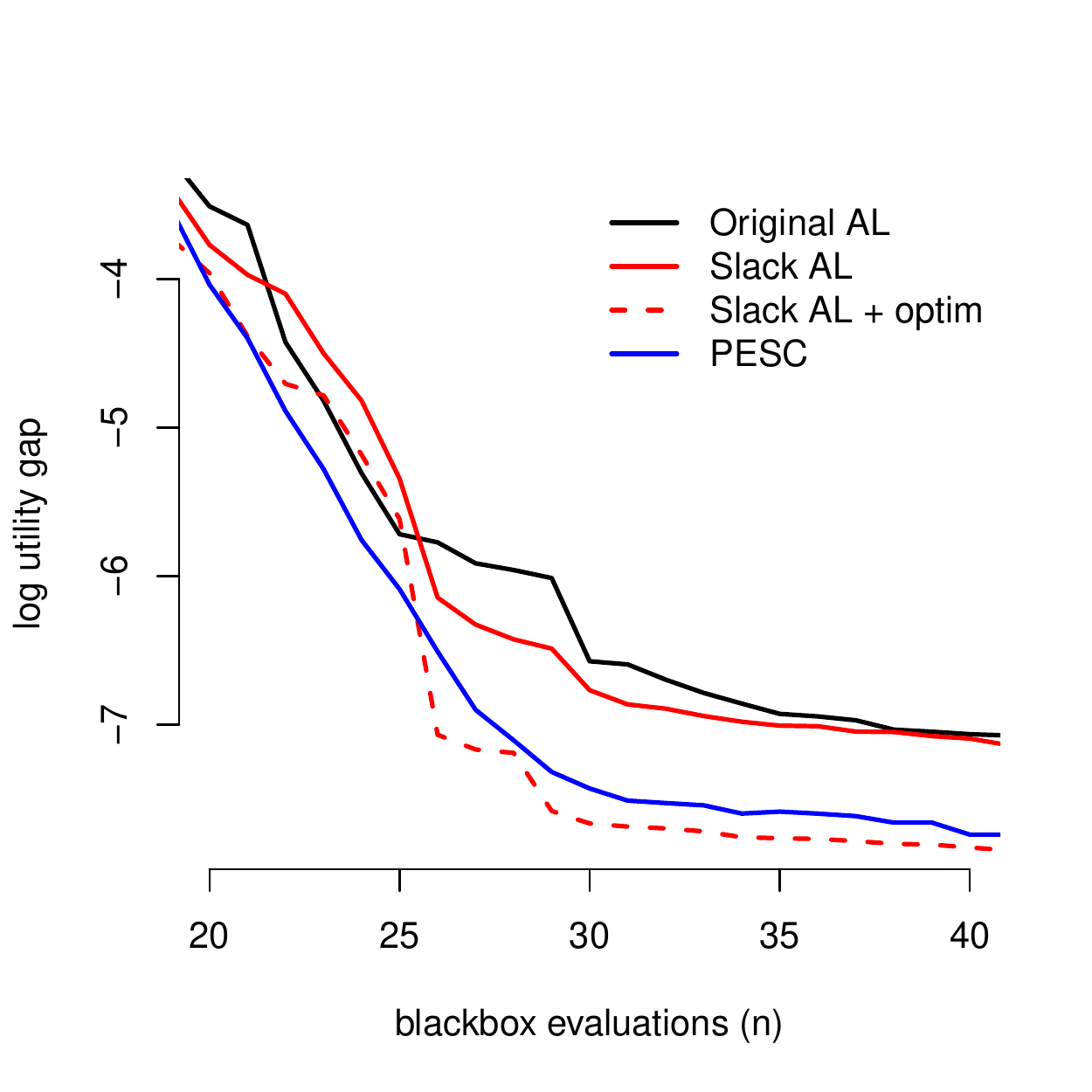}
\caption{Results on the LSQ problem with initial designs
of size $n=10$.  The {\em left} panel shows the best valid value of the
objective over the first 40 evaluations, whereas the right shows the $\log$
{\em utility-gap} for the second 20 evaluations. The solid gray lines show
comparators from~\cite{GrGrLedLeeRaWeWi2016}.}
\label{f:toy}
\end{figure}
The {\em left}-hand plot in Figure~\ref{f:toy} tracks the average best valid 
value of 
the objective found over the iterations, using the progress metric described
above.  Random initial designs of size $n=5$ were used, as indicated by the
vertical-dashed gray line.  The solid {\em gray} lines are extracted from a
similar plot from~\cite{GrGrLedLeeRaWeWi2016}, containing both AL-based comparators, and
several from the derivative-free optimization and BO literatures.  The details
are omitted here.
%, except we note that
%those gray AL comparators did not utilize the initialization and updating
%scheme described by~\citet{rejoinder}, and which we updated for mixed
%constraints in Section
%\ref{sec:alg}.  
Our new ALBO comparators are shown in thicker colored lines; the solid black
line is the original AL(BO)-EI comparator, under a revised (compared to~\citep{GrGrLedLeeRaWeWi2016}) initialization and updating scheme.  The two red
lines are variations on the slack-AL algorithm under EI: with (dashed) and
without (solid) L-BFGS-B optimizing EI acquisition at each iteration. Finally,
the blue line is PESC~\cite{hernandez:etal:2015}, using the {\sf Python}
library available at \url{https://github.com/HIPS/Spearmint/tree/PESC}.
%\footnote{About 15\% of the
%PESC runs encountered a runtime error; these were omitted from the
%presentation without checking for any potential bias.} 
The take-home message
from the plot is that all four new methods outperform those considered by the
original ALBO paper~\cite{GrGrLedLeeRaWeWi2016}. 

Focusing on the new comparators only, observe that their progress is {\em
nearly} statistically equivalent during the first 20 iterations. At iteration
10, for example, the ordering from best to worst is PESC (0.861), Slack-AL
(0.902), Orig-AL (0.942), and Slack-AL+optim (0.957). Combining the 100
repetitions of each run, only the best (PESC) versus worst (Slack-AL+optim) is
statistically significant in a one-sided $t$-test, with a $p$-value of 0.0011.
Apparently, aggressively optimizing the EI in the slack formulation in early
iterations hurts more than it helps.  The situation is more nuanced, however,
for later iterations. At iteration 30, for example, the ordering is
Slack-AL+optim (0.6002), PESC (0.6004), Slack-AL (0.6010), and Orig-AL
(0.689).  Although the numbers are more tightly grouped, only the first two
(Slack-AL+optim and PESC), both leveraging L-BFGS-B subroutines, are
statistically equivalent. In particular, both Slack-AL variants outperform the
original AL with $p$-values less than $2.2e^{-16}$.  This discrepancy is more
easily visualized in the {\em left} of the figure with a so-called
``utility-gap'' log plot~\citep{hernandez:etal:2015}, which tracks the log difference between the
theoretical best valid value and those found by search.

\subsection{Mixed inequality and equality constrained problems}
\label{sec:mixed}

Next consider a problem in four input dimensions with a (known) linear
objective and two constraints, one inequality and one equality. The inequality
constraint is the so-called ``Ackley'' function in $d=4$ input dimensions. %\citep[for details see][]{adorio:diliman:2013,molga:smutnicki:2005,back:1996}.
%\citet{simulationlib} provide a convenient reference along with {\sf R} and {\sf
%MATLAB} implementations.  Following the description on that page, we used the
%recommended values of $a = 20$, $b = 0.2$, and $c = 2\pi$.
%
The equality constraint follows the so-called ``Hartman 4-dimensional function''.
%\citep[see][for details]{dixon:szego:1978,picheny:wagner:ginsbourger:2012}.
%Again~\citet{simulationlib} is a convenient reference with code and
%visualizations. Both ``Hartman-4'' and ``Ackley'' functions have been used to
%illustrate {\em unconstrained} blackbox optimization algorithms.
% SW does not believe this:
% we believe that ours is the first use as constraints in blackbox optimization. 
Appendix~\ref{sec:testpbs}
provides a full mathematical specification.
\begin{figure}[ht!]
\centering
%\vspace{-0.4cm}
\includegraphics[scale=0.625,trim=0 20 0 40]{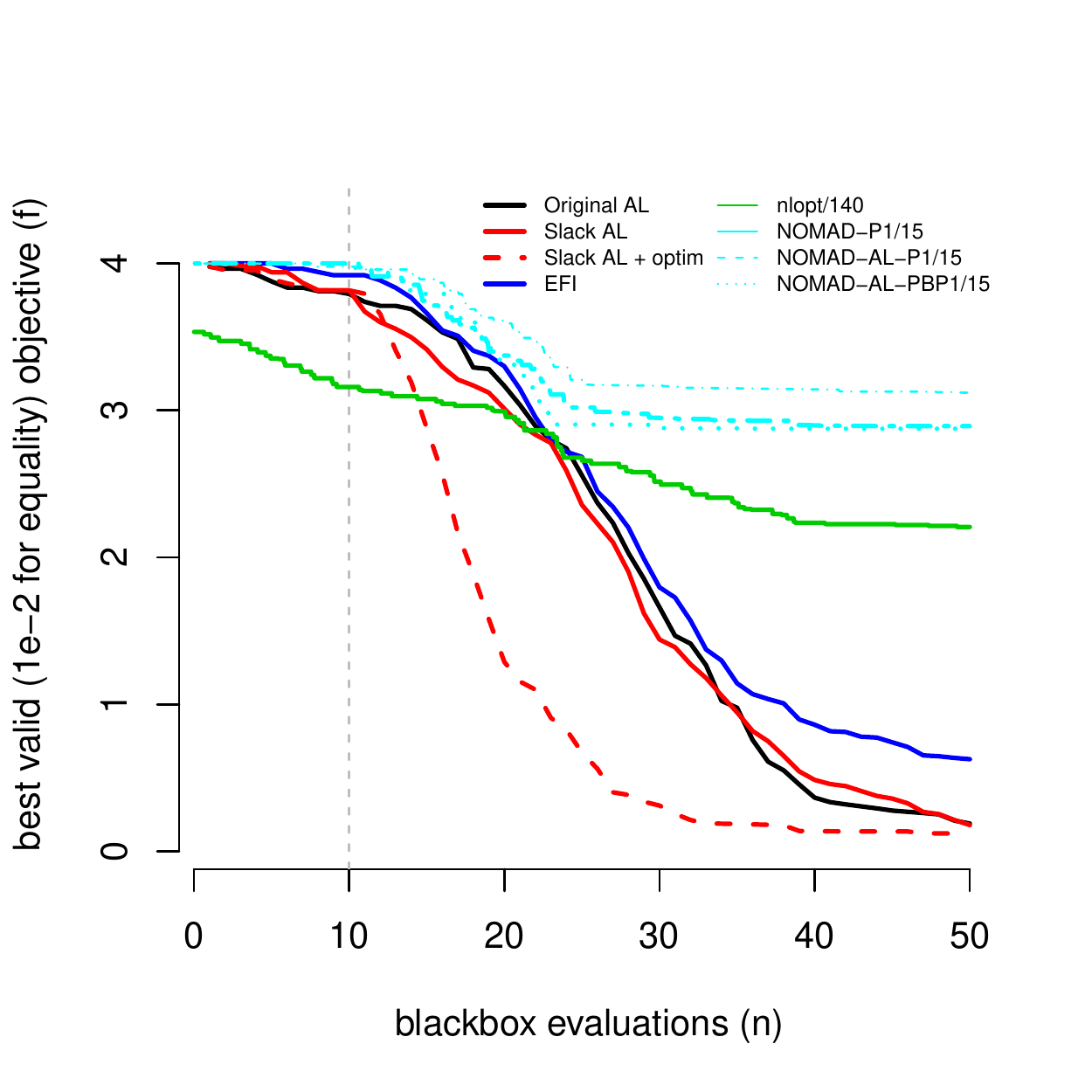} \hfill
\includegraphics[scale=0.625,trim=0 20 0 40]{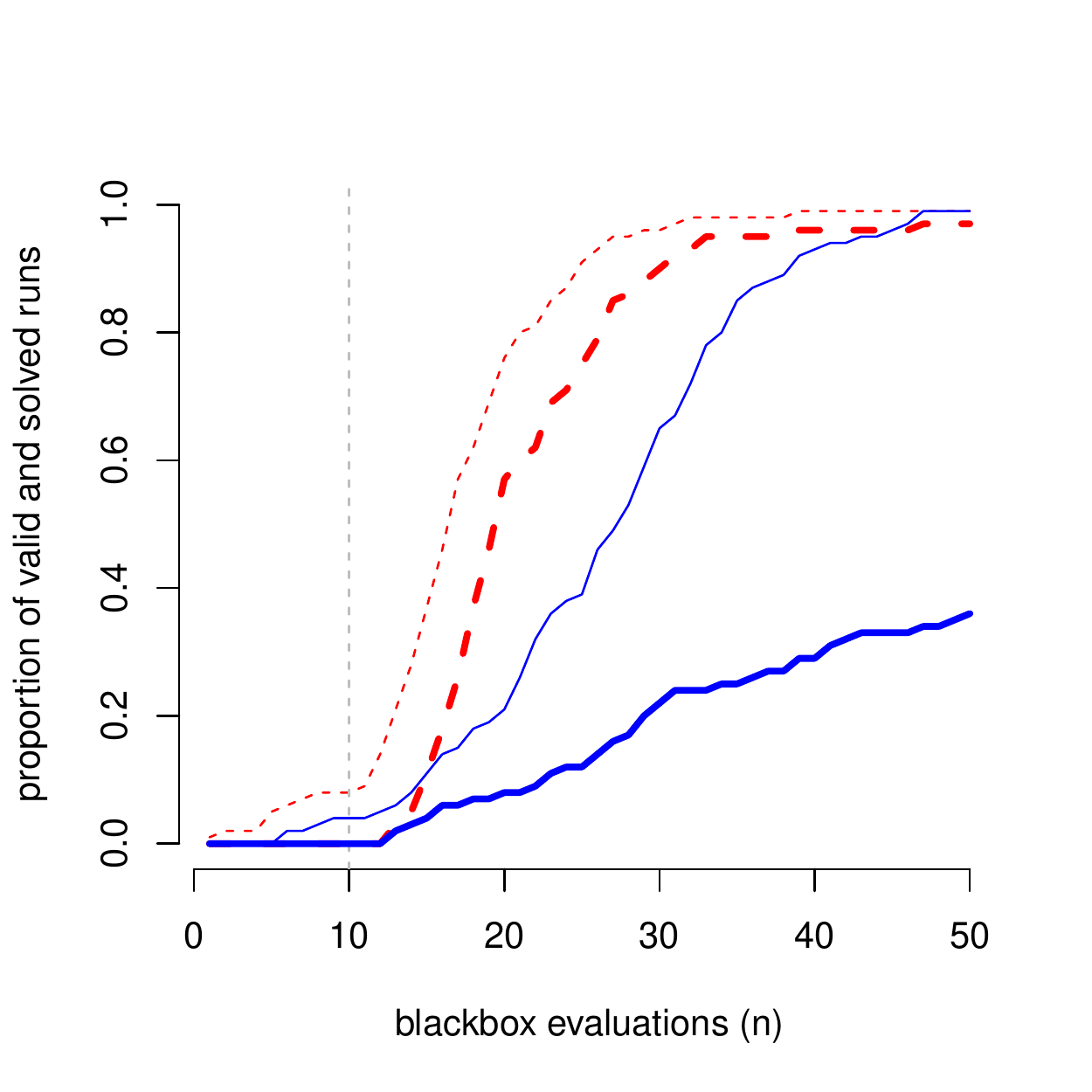} %\hspace{0.25cm}
% \vspace{-0.5cm}
% \includegraphics[width=.305\textwidth,trim=27 0 3 40]{pb18-1}
% \includegraphics[width=.305\textwidth,trim=10 0 20 40]{pb18-2}
\caption{Results on the Linear-Hartman-Ackley mixed constraint problem.  The {\em left} panel shows
a progress comparison based on {\tt laGP} code with initial designs of size
$n=10$. The $x$-scale has been divided by 140 for the {\sf nlopt} comparator.
A value of four indicates that no valid solution has been found. The {\em
right} panel shows the proportion of valid (thin lines) and optimal (thick
lines) solutions for the EFI and ``Slack AL + optim'' comparators. }
\label{f:mixed}
\end{figure}
Figure~\ref{f:mixed} shows two views into progress on this problem. Since
it involves mixed constraints, comparators from the BO literature are scarce.
Our EFI implementation deploys the $(-h,h)$ heuristic mentioned in the
introduction. As representatives from the nonlinear optimization literature we
include {\sf nlopt}~\citep{nlopt} and three adapted {\sf NOMAD}~\citep{Le09b}
comparators, which are detailed in Appendix~\ref{sec:nomad}. In the {\em
left}-hand plot we can see that our new ALBO comparators are the clear winner,
with an L-BFGS-B optimized EI search under the slack-variable AL
implementation performing exceptionally well. The {\sf nlopt} and {\sf NOMAD}
comparators are particularly poor.  We allowed those to run up to 7000 and 1000
iterations, respectively, and in the plot we scaled the $x$-axis (i.e., $n$)
to put them on the same scale as the others. The {\em right}-hand plot
provides a view into the distribution of two key aspects of performance over
the MC repetitions. Observe that ``Slack AL + optim'' finds valid values
quickly, and optimal values not much later. Our adapted EFI is particularly
slow at converging to optimal (valid) solutions.

Our final problem involves two input dimensions, an {\em unknown} objective
function (i.e., one that must be modeled with a GP), one inequality constraint
and two equality constraints.
% : $\min_{x \in [0,1]^2} f_2(x)$, subject to $c_3(x)
% \leq 0$, and $c_{4,5}(x) = 0$
The objective is a centered and re-scaled version of the ``Goldstein--Price''
function.
%\citep{dixon:szego:1978,molga:smutnicki:2005,picheny:wagner:ginsbourger:2012}.
The inequality constraint is the sinusoidal constraint from the LSQ
problem [Section~\ref{sec:ineq}]. The first equality constraint is a centered
``Branin'' function, %\citep{dixon:szego:1978,forrester:sobester:keane:2008,molga:smutnicki:2005,picheny:wagner:ginsbourger:2012};
the second equality constraint is taken from~\cite{parr:etal:2012}
(henceforth GBSP problem).
% I don't think this is worth bragging about, it just looks like you went 
% % hunting for problems that you looked good on:
% As before, all four
% functions have previously been used in a blackbox optimization context;
% however ours is the first of its kind in a mixed constraint context. 
Appendix~\ref{sec:testpbs} contains a full mathematical specification.
%  and
% visualization. We shall refer to this as the ``GGBP'' problem, combining
% initials from the names and/or authors of the corresponding functions. Further
% description and code is provided by~\citet{bingham:etal:2014}.
\begin{figure}[ht!]
\centering
%\vspace{-0.4cm}
\includegraphics[scale=0.625,trim=0 20 0 40]{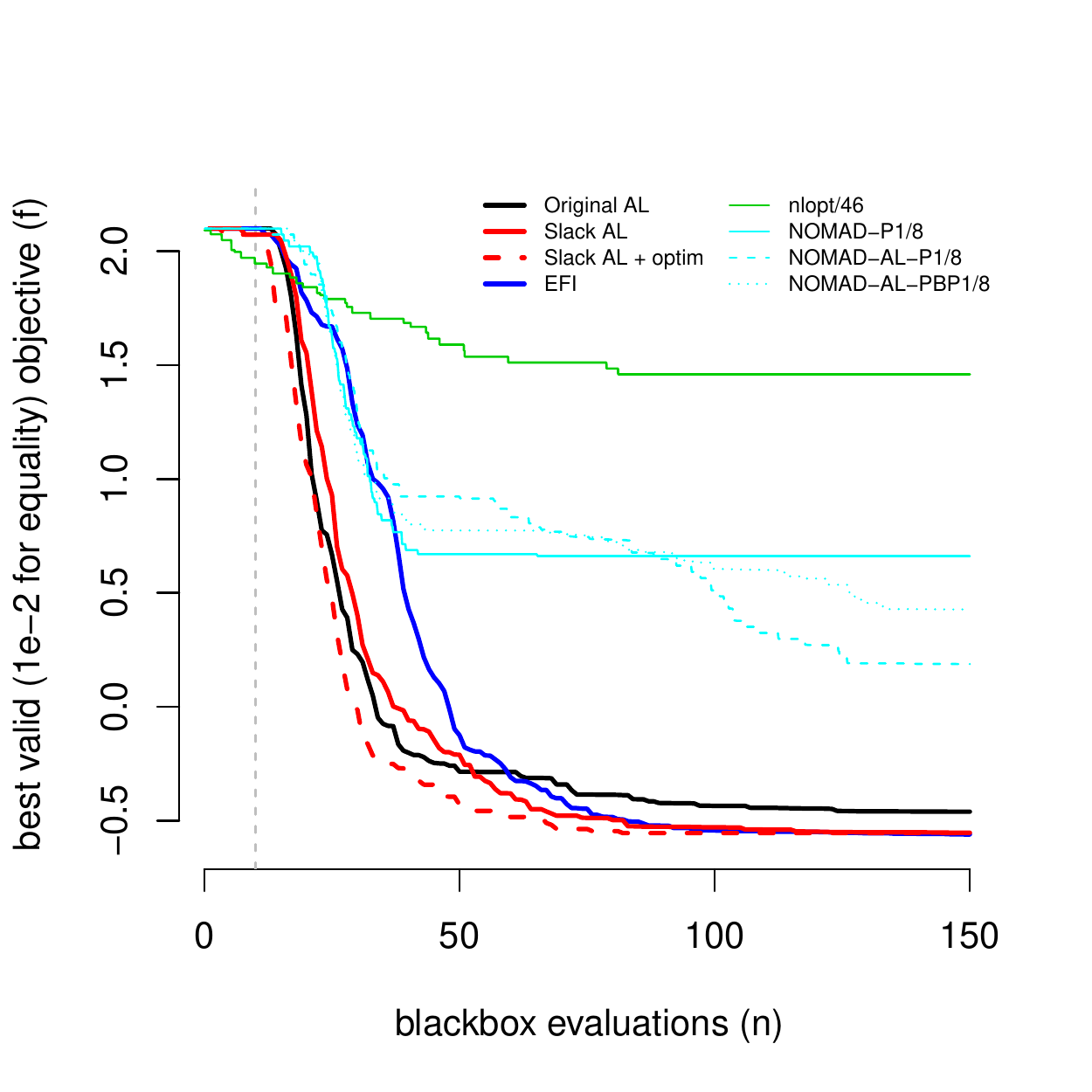} \hfill %\hspace{0.25cm}
\includegraphics[scale=0.625,trim=0 20 0 40]{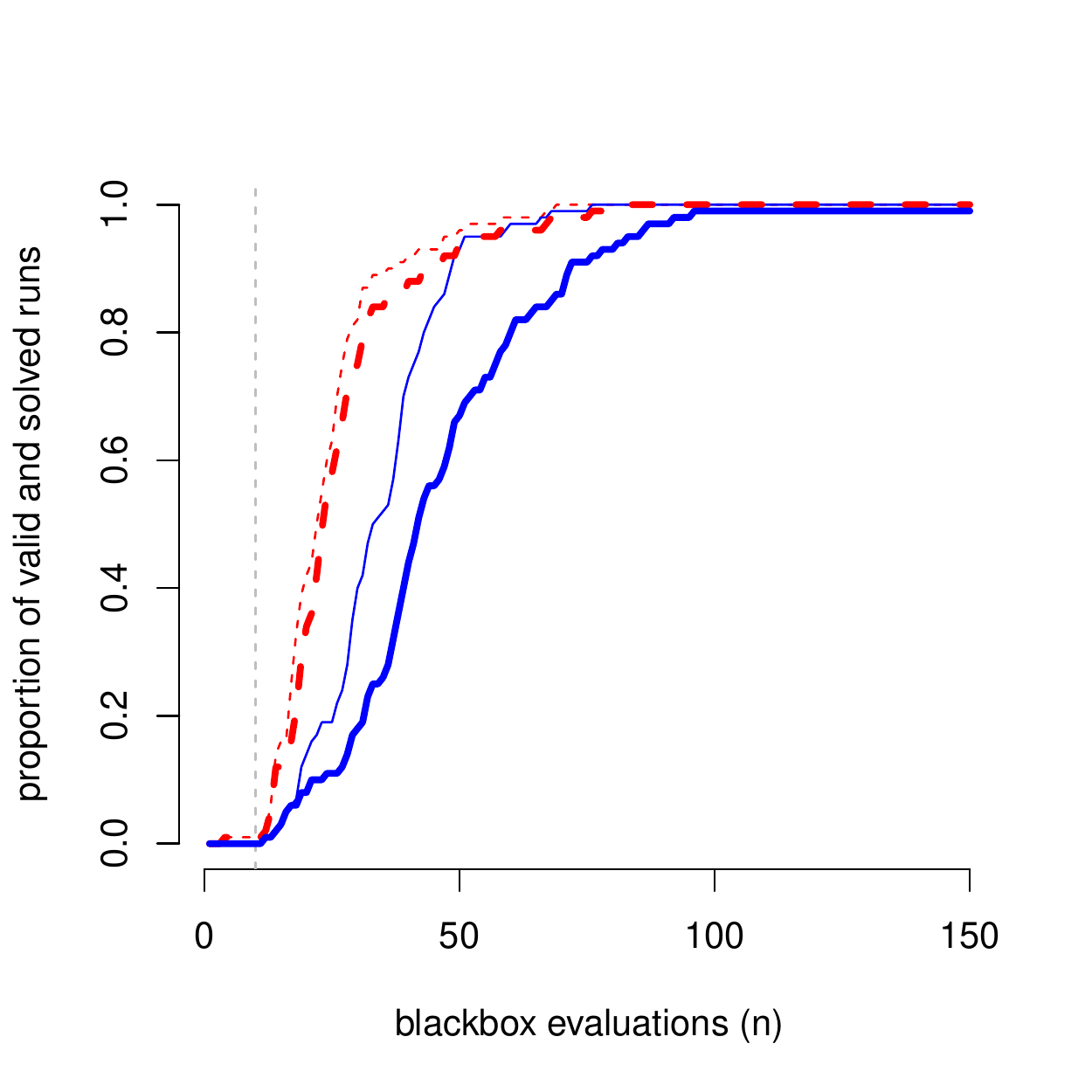} %\hspace{0.25cm}
\caption{Results on the GBSP problem.  
See Figure~\ref{f:mixed} caption.}
\label{f:mixed13}
\end{figure}
Figure~\ref{f:mixed13} shows our results on this problem.  
% The {\em left} panel tracks the average best valid value as in previous experiments. 
Observe ({\em left} panel)
that the original ALBO comparator makes rapid progress at first, but dramatically
slows for later iterations.  The other ALBO comparators, including EFI,
converge much more reliably, with the ``Slack AL + optim'' comparator leading
in both stages (early progress and ultimate convergence).  Again, {\sf nlopt}
and {\sf NOMAD} are poor, however note that their relative comparison is
reversed; again, we scaled the $x$-axis to view these on a similar scale as
the others. The {\em right} panel shows the proportion of valid and optimal
solutions for ``Slack AL + optim'' and EFI. Notice that the AL method finds an
optimal solution almost as quickly as it finds a valid one---both
substantially faster than EFI.

\section{Discussion}
\label{sec:discuss}

The augmented Lagrangian (AL) is an established apparatus from the
mathematical optimization literature, enabling unconstrained or 
bound-constrained optimizers to be
deployed in settings with constraints.  Recent work involving Bayesian
optimization within the AL framework (ALBO) has shown great promise,
especially toward obtaining global solutions under constraints. However, those
methods were deficient in at least two respects.  One is that only
inequality constraints could be supported.  Another was that evaluating the
acquisition function, combining predictive mean and variance information via
EI, required Monte Carlo approximation.  In this paper
we showed that both drawbacks could be addressed via a slack-variable
reformulation of the AL.  Our method supports inequality, equality, and mixed
constraints, and to our knowledge this updated ALBO procedure is unique in
the BO literature in its applicability to the general mixed constraints
problem~\eqref{eq:ineqprob}.  We showed that the slack ALBO method
outperforms modern alternatives in several challenging constrained
optimization problems.

We conclude by remarking on a potential drawback: we have found in some cases
that our slack variable approach is a double-edged sword, especially when the
unknown slacks are chosen in a default manner, i.e., $s^*(x)$ as in Section~\ref{sec:sopt}.  Those choices utilize the surrogate(s) of the constraints,
particularly the posterior mean, which means that those surrogates are relied
upon more heavily than in the original (non-slack) AL context.  Consequently,
if the problem at hand matches the surrogate modeling (i.e., Gaussian process)
assumptions well, then by a leveraging argument we can expect the slack method
to do better than the original. However, if the assumptions are a mismatch,
then one can expect them to be worse.   The motivating ``Lockwood'' problem
from~\citet{GrGrLedLeeRaWeWi2016} had a constraint surface that was kinked at
its boundary $\{x\in \R^d : c(x) = 0\}$, which obviously violates the 
smoothness assumptions
of typical GP surrogates.  We were therefore not surprised to find poorer
performance in our slack method compared to the original.

\subsection*{Acknowledgments}

We are grateful to Mickael Binois for comments on early drafts.  RBG is
grateful for partial support from National Science Foundation grant
DMS-1521702. 
The work of SMW is supported by the U.S.\ Department of Energy, Office of 
Science, Office of Advanced Scientific Computing Research under Contract No.\ 
 DE-AC02-06CH11357.  
The work of SLD is supported by the Natural Sciences and Engineering Research Council of Canada
grant 418250.

\appendix

\section{Optimal and default slack variables}
\label{sec:optslack}

This continues the discussion from Section~\ref{sec:alg}, which suggests
choosing slack variables by minimizing the slack-AL.  One alternative is to
maximize the EI. Recall that EI mixes data quantities (at the $x_i$) and
predictive quantities at candidate locations $(x,s)$ via the improvement
$I(x,y) =
\max(0, y_{\min}^n - Y(x,s))$, the former coming from $y_{\min}^n = \min(y_1,
\dots, y_n)$, where $y_i = y(x_i, s)$.  If one seeks improvement
over the lowest possible $y_{\min}^n$, then the optimal $s$-values to pair
with the $x_i$s are given in Eq.~(\ref{eq:sopt}):
use $y_i = y(x_i, s^*(x_i))$. Choosing slacks for candidate $(x,s)$ locations
is more challenging. Solving  $\max_{s \in \Rp^m} \mathrm{EI}(x,s)$ for a
fixed $x$ is equivalent to solving  $\max_{s
\in \Rp^m} \mathbb{E}\{ Y(x,s) \mathbb{I}_{Y(x,s)
\leq y_{\min}}\}$, for which a closed-form solution remains elusive.
Numerical methods are an option; however, we have not discovered any
advantage over the simpler $\min_{s \in \Rp^m} \mathbb{E}
\{ Y(x,s)
\}$-based settings described above.  Empirically, we find that the
two criteria either yield identical $s^*_j(x)$-values, or ones that are nearly
so.

As an illustration, we consider a 2-d problem with a linear objective and two
inequality constraints (see the LSQ problem in Section~\ref{sec:testpbs}, Eq.~(\ref{eq:gramacypb}), below). We focus here on a ``static''
situation, where a 9-point grid defines the set of observations. $f_1$ is
treated as a known function and two GP models are fitted to $c_1$ and $c_2$.
The AL parameters are set to $\rho=1/16$ and $\lambda= [1,1]$.
\begin{figure}[th!]
\centering
\includegraphics[scale=0.8]{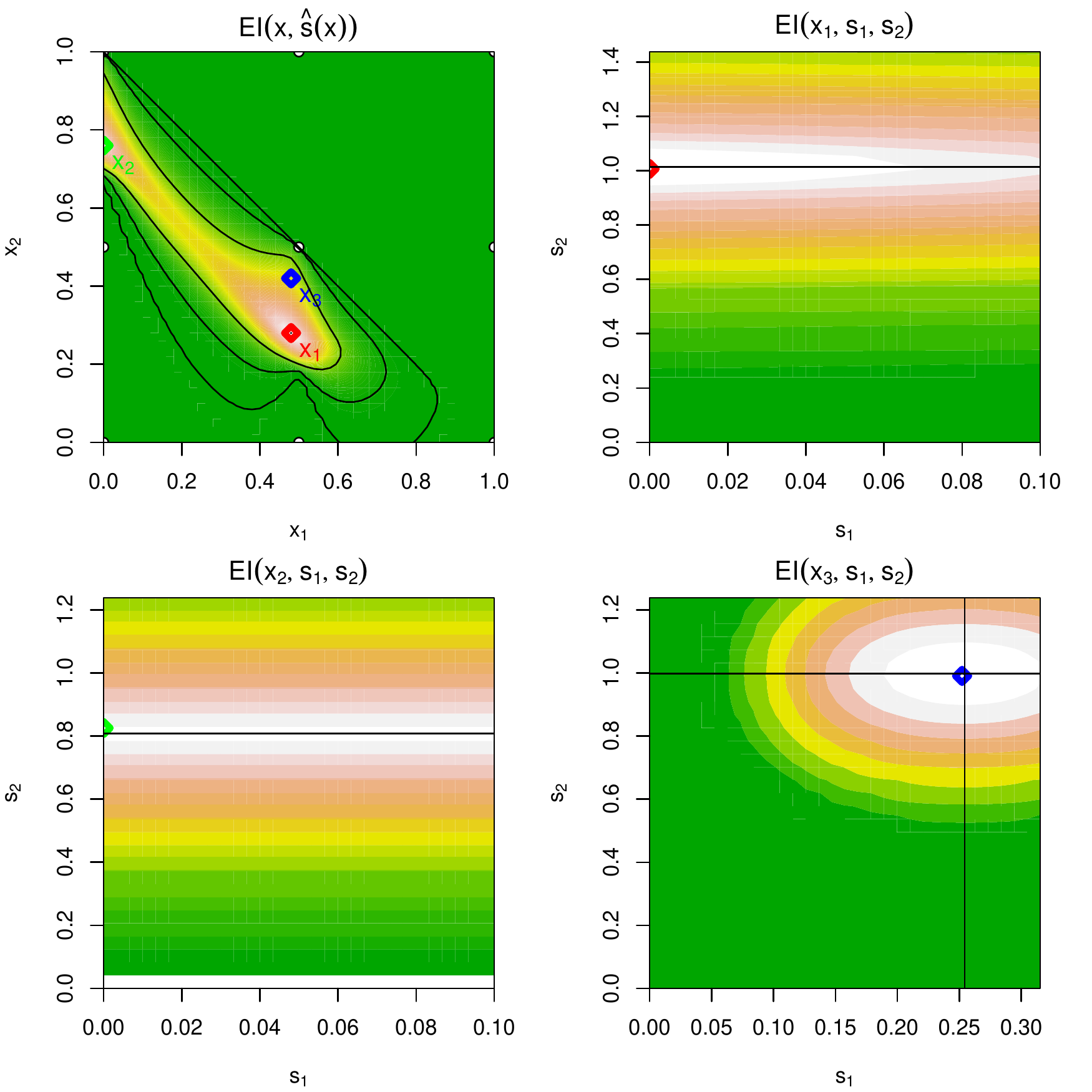} %\hspace{0.5cm}
\caption{Illustration of the influence of the slack variables. The top left figure shows  
EI  as a function of $x$, when the slack variables are chosen as $\hat
s$. The three other figures show the EI metric as a function of the slack
variables for three different $x$-values (at locations shown in the top left
graph). The horizontal and vertical lines show the $s^*$-values, and the
points show the optimal values.}
\label{f:slackExplained}
\end{figure}
The {\em top-left} panel of Figure~\ref{f:slackExplained} shows the EI  
surface (computed on a 2-d grid), when $s_j^*(x) = \max
\left( 0, - \lambda_j\rho - \mu_j(x) \right)$ is used for each grid element 
$x$. The shape of the contours is typical of EI, with a plateau of zero
values and two local maxima. The remaining panels show the
value of the EI metric for each of the three $x$-values shown in the {\em 
top-left} panel. These are computed on a 2-d
grid in $s$ space. Despite its complex formulation (\ref{eq:EIs}), on this 
example EI is
apparently unimodal with respect to $s_1$ and $s_2$. Moreover, the optimal
values (computed from the grid used to plot the image) coincide (up to
numerical error) with the associated $s^*$ values,
plotted as diamonds matching the color in the {\em top-left} plot.

\section{Implementation notes}

\subsection{Initializing the AL parameters in the slack setup}
\label{sec:alparams}

%% SUPMAT
To initialize $(\lambda^0, \rho^0)$, we update the settings suggested in~\cite{rejoinder}---balancing the scales of objective and constraint in the AL
on the initial design of size $n_0$---to our new mixed constraint context. Let
$v(x)$ be a logical vector of length $m+p$ recording the validity of $x$ in a
zero-slack setting.  That is, let $v_j(x) = 1$ if $c_j(x) \leq 0$, for
$j=1,\dots,p$, and $v_j(x) = 1$ if $|c_j(x)| \leq \epsilon$, for 
%! SW: Note I chanced ``< \epsilon'' to ``\leq epsilon'' above
$j=p+1,\dots,m$;
otherwise let $v_j(x) = 0$.
Then, take
\[
\rho^0 = \frac{\min_{i=1,\dots, n_0} \{ \sum_{j=1}^{m+p}
c_j(x_i)^2 : \exists j, v_j(x_i) = 0 \; \}
}{2 \min_{i=1, \dots, n_0}\{f(x_i) : \forall j,  v_j(x_i) = 1\}},
\]
and $\lambda^0 = 0$. The denominator above is not defined if the initial
design has no valid values (i.e., if there is no $x_i$ with $v_j(x_i) = 1$
for all $j$). When that happens we use the median of $f(x_i)$ in the
denominator instead. On the other hand, if the initial design has no invalid 
values and hence the numerator is not defined, we use $\rho^0=1$.

\subsection{Open source software}
\label{sec:open}

Code supporting all methods described here is provided in two open source {\sf
R} packages: {\tt laGP}~\citep{laGP} and {\tt DiceOptim}~\citep{DiceOptim}
packages, both on CRAN. Implementation details vary somewhat
across those packages, due primarily to particulars of their surrogate
modeling capability and how they search the EI surface.  For example {\tt
laGP} can accommodate a smaller initial sample size because it learns fewer
parameters (i.e., has fewer degrees of freedom). {\tt DiceOptim} uses a
multi-start search procedure for EI, whereas {\tt laGP} deploys a random
candidate grid, which may optionally be ``finished'' with an L-BFGS-B search.
Nevertheless, their qualitative behavior exhibits strong similarity. Both
utilize library subroutines for dealing with weighted non-central chi-square
deviates, as described in Appendix~\ref{sec:wsnc} below. Some notable
similarities and differences are described below. The empirical work we
present in Section~\ref{sec:empirical} provides results from our {\tt laGP}
only.  Those obtained from {\tt DiceOptim} are similar; we encourage readers
to consult the documentation from both packages for further
illustration.\footnote{At the time of writing, updated versions of {\tt laGP}
and {\tt DiceOptim} are not ``pushed'' to CRAN; alpha testing versions are available from the authors upon
request.  Subsequent CRAN updates will include all slack AL enhancements made
for this paper.}

Both  initialize $\rho$ and $\lambda$ as described above in Section~\ref{sec:alparams} above; both update those parameters according to the
modifications of steps 2--4 in Alg.~\ref{alg:baseal}; and both define
$x^k$ in those updates to be the value of $x_i$ corresponding to the best $y_i
= y(x_i, s^*(x_i))$ using $(\lambda^{k-1},
\rho^{k-1})$, for data values indexed by $i=1,\dots,n$.  By default, both
utilize $(x,s^*(x))$ to evaluate EI at candidate $x$-locations, via the
surrogate mean $\mu_c(x)$ values.  The {\tt laGP} implementation optionally
allows randomization of $(x,S^*(x))$, where $S^*(x)$ is chosen via (14) with
a draw $Y_{c_j}(x)$ in place of $c_j(x)$. Alternatively, {\tt DiceKriging} can
optimize jointly over $(x,s)$ via its multi-start search scheme.  Those two
options are variations which allow one to test the ``optimality'' of $s^*(x)$
under EI calculations. We report here that such approaches have been found to
be uniformly inferior to using $s^*(x)$, which is anyways a much simpler
implementation choice.

Choosing $s^*(x)$, as opposed to random $S^*(x)$ above, renders the entire EI
calculation for a candidate $x$ value deterministic.  In turn, that means that
local (numerical) derivative-based optimization techniques, such as
L-BFGS-B~\citep{liu:nocedal:1989}, can be used to optimize over EI settings.
The {\tt laGP} package has an optional ``{\tt optim}'' setting, initializing
local L-BFGS-B searches from the best random candidate point via.  {\tt
DiceKriging} uses the \emph{genoud} algorithm~\citep[GENetic Optimization
Using Derivatives,][]{mebane:sekhon:2011}to handle potentially multi-modal EI
surfaces (Figure~\ref{f:slackExplained}). Neither of these options is
available for the original AL (without slacks), since evaluating the EI
required Monte Carlo, effectively limiting the resolution of search.  In later
iterations, the original AL would fail to ``drill'' down into local troughs of
the EI surface. In our empirical work [Section~\ref{sec:empirical}], we show
that being able to optimize over the EI surface is crucial to obtaining good
performance at later iterations.  The {\em predictive entropy search}
adaptation for constrained optimization~\citep[PESC;][]{hernandez:etal:2015}
is similarly deterministic, and also utilizes L-BFGS-B to optimize over
selections.  We conjecture that the correspondingly suboptimal selections made
by the original AL, and the poor initialization of updates of $\rho$ pointed
out by~\cite{discussion4}, explains the AL-versus-PESC outcome reported
by~\cite{hernandez:etal:2015}. Our revised experiments show a reversed (if
extremely close) comparison.

Finally, we find that the following simple-yet-very-efficient numerical trick
can be helpful when the EI surface is mostly zero, as often happens at later
stages of optimization---a behavior that can be attributed to the quadratic
nature of the AL, guaranteeing zero improvement in certain
situations~\citep{GrGrLedLeeRaWeWi2016}. In the original {\tt laGP}
implementation this drawback was addressed by ``switching'' to mean-based
search (rather than EI). In the slack variable formulation, observe that zero
EI may be realized when $w^n_{\min} < 0$, in which case the $w^n_{\min}$ value
itself (as it is less than zero) may stand in as a sensible replacement:
capturing both mean and (deterministic aspects) of the penalty information.
For example, when optimizing over EI in {\tt DiceOptim} we find that this
$w^n_{\min}$ replacement helps the solver escape large plateaus of (otherwise)
zero values.

\subsection{Adapted NOMAD comparators}
\label{sec:nomad}

The three methods {\tt NOMAD-P1}, {\tt NOMAD-AL-P1}, and {\tt NOMAD-AL-PB-P1}
are based on the {\sf NOMAD} software~\cite{Le09b} that implements the Mesh
Adaptive Direct Search derivative-free algorithm~\cite{AuDe2006} using the
Progressive Barrier (PB) technique~\cite{AuDe09a} for inequality constraints.
All three methods begin with a first phase ({\tt P1}) that focuses on
obtaining a feasible solution, with the execution of {\sf NOMAD} on the
minimization of $\sum_{k=1}^p h_k^2(x)$ subject to $g_j(x) \leq 0, j =
1,\ldots,m$.
Then, {\tt NOMAD-P1} follows with an ordinary {\sf NOMAD} run where equalities
are transformed into  inequalities of the form $|h_k(x)| \leq 0, k=1, \ldots,
p$.
{\tt NOMAD-AL-P1} is a straightforward extension of the augmented Lagrangian
method of~\cite{GrGrLedLeeRaWeWi2016}, with the inclusion of equality
constraints in the Lagrangian function, while the unconstrained subproblem is
handled by {\sf NOMAD}.
{\tt NOMAD-AL-PB-P1} is a hybrid version where equalities are still in the
Lagrangian while inequalities are transferred into the subproblem where {\sf
NOMAD} treats them with the PB.
Note that the non-use of the first phase 
{\tt P1} was also tested, but it did not work as well as the others so it was
not included in order to save space.

\section{Code for calculations with WSNC RVs}
\label{sec:wsnc}

Here we provide {\sf R} code for calculating EI and related quantities using
the {\tt sadists}~\citep{sadists-Manual}  and {\tt CompQuadForm}~\citep{CompQuadForm-Manual} libraries, in particular their WNCS distribution
functions~\citep[e.g.,][]{davies:1980,duchense:lafaye:2010}.

Let {\tt mz} and {\tt sz} denote the vectorized
means and variances of the normal $Z_i$ quantity in (\ref{eq:Z}), that is
\verb!mx[j]! $=\mu_{c_j}(x) + \alpha_j$. Below, {\tt rho} contains the AL
parameter $\rho$, and  we presume $f(x)$, $r(s)$, and $y^n_{\min}$ quantities
have been pre-calculated and stored in variables of the same name.  Then, the
relevant quantities for calculating EI in the case of a known $f(x)$ can be
computed as:

\begin{verbatim}
R> ncp <- (mz / sz)^2
R> wmin <- 2*rho*(ymin - fx - gs) 
\end{verbatim}
Using {\tt sadists}, EI may be computed as follows using simple ``quadrature''.

\begin{verbatim}
R> library(sadists)
R> m <- length(mz)
R> lt <- 1000
R> t <- seq(0, wmin, length=lt)
R> df <- pow <- rep(1, mz)
R> ncp <- (mx / sx)^2
R> EI <- (sum(psumchisqpow(q=t, sz^2, df, ncp, pow))*wmin/(lt-1)) / (2*rho)
\end{verbatim}
Alternately, via the {\tt integrate} function built-into {\sf R}:
\begin{verbatim}
R> EIgrand <- function(t) { psumchisqpow(q=t, wts, df, ncp, pow) }
R> EI <- integrate(EIgrand, 0, wmin)$value / (2*rho)
\end{verbatim}

The code using {\tt CompQuadForm} is similar.  Greater care is required to code
the EI integrand, as the main function ({\tt davies}) is not vectorized.  Also,
we have found that {\tt NaN} is often erroneously returned when the integrand
is actually zero.

\begin{verbatim}
library(CompQuadForm)
R> EIgrand <- function(x) { 
+     p <- rep(NA, length(x))
+     for(1 in 1:length(c))  
+       p[i] <- 1 - davies(q=x[i], lambda=sz^2, delta=ncp)$Qq
+     p[!is.finite(p)] <- 0
+     return(p)
}
R> EI <- (sum(EIgrand(t)*wmin/(lt-1)) / (2*rho)
R> EI <- integrate(EIgrand, 0, wmin)$value / (2*rho) ## alternately
\end{verbatim}
Although this may at first seem more cumbersome, it is actually fairly
easy to vectorize {\tt davies}, and at the same time replace {\tt NaN} values
with zero in the underlying {\sf C} implementation.  The result is a method
which is much faster than the {\tt sadists} alternative, and has the added
bonus of being callable from the {\sf C} functions inside our {\tt laGP}
implementation.  For more details on this re-implementation of {\tt davies} in
our setup, please see the source for the {\tt laGP} and/or {\tt
DiceOptim} package(s).

Some slight modification is required for the unknown objective case. Let {\tt
mf} and {\tt sf} denote $\mu_f(x)$ and $\sigma_f(x)$ in the code below. We
have concluded that only the methods from {\tt CompQuadForm} are applicable in
this case.
\begin{verbatim}
R> alpha <- 2*rho*lambda + 2*s
R> wmin <- 2*rho*ymin - sum(s^2) - s %*% t(lambda) + sum(alpha^2/4)
t <- seq(-6*rho*sf, wmin, length=lt)
R> EIgrand <- function(x) { 
+     p <- rep(NA, length(x))
+     madj <- 2*rho*mf
+     sadj <- 2*rho*sf
+     for(1 in 1:length(c))  
+       p[i] <- davies(q=x[i]-madj, lambda=sz^2, delta=ncp, sigma=sadj)$Qq
+     p <- 1-p
+     p[!is.finite(p)] <- 0
+     return(p)
}
R> EI <- (sum(EIgrand(t))*(wmin+6*rho*s)/(lt-1)) / (2*rho)
R> EI <- integrate(EIgrand, -6*rho*sf, wmin)$value / (2*rho) ## alternately
\end{verbatim}
As when $f(x)$ is known, vectorizing the {\tt davies} routine leads
to dramatic speedups.

\section{Test problems}
\label{sec:testpbs}

Below we detail the components of the three test problems explored in Section~\ref{sec:empirical}.
They combine classic benchmarks from the Bayesian
optimization and computer (surrogate) modeling literature.  In total there are
two objective functions and six constraints.

\medskip
\noindent{\bf Objective functions}:
$f_1$ is a simple linear objective, treated as known, and 
$f_2$ is the \textit{Goldstein-Price} function (rescaled and centered)
\cite{dixon:szego:1978,molga:smutnicki:2005,picheny:wagner:ginsbourger:2012}.
Further description, as well as {\sf R} and {\sf MATLAB} implementations,  is
provided by~\cite{bingham:etal:2014}.

\vspace{-0.5cm}
\begin{align*}
f_1(x) &= \sum_{i=1}^d x_i \\
f_2(x) &= \frac{\log \left[(1+a) (30 + b) \right] - 8.69}{2.43}, \quad \mbox{with} \\
    a &= \left(4 x_1 + 4 x_2 - 3 \right)^2  \times \\
    & \left[75 - 56 \left(x_1 + x_2 \right) + 3\left(4 x_1 - 2 \right)^2 + 6\left(4 x_1 - 2 \right)\left(4 x_2 - 2 \right) + 3\left(4 x_2 - 2 \right)^2\right]\\
b &= \left(8 x_1 - 12 x_2 +2 \right)^2 \times \\
& \left[-14 - 128 x_1 + 12\left(4 x_1 - 2 \right)^2 + 192 x_2 - 36\left(4 x_1 - 2 \right)\left(4 x_2 - 2 \right) + 27\left(4 x_2 - 2 \right)^2 \right]
\end{align*}

\medskip
\noindent {\bf Constraint functions}:
\noindent 
$c_1$ and $c_2$ are the ``toy'' constraints from~\cite{GrGrLedLeeRaWeWi2016}; 
$c_3$ is the ``Branin'' function (centered) ``Branin'' function~\citep{dixon:szego:1978,forrester:sobester:keane:2008,molga:smutnicki:2005,picheny:wagner:ginsbourger:2012};
$c_4$ is taken from~\cite{parr:etal:2012}.
$c_5$ is the ``Ackley'' function (centered)~\citep[for details see][]{adorio:diliman:2013,molga:smutnicki:2005,back:1996};
$c_6$ the ``Hartman'' function (centered, rescaled)~\citep[see][for details]{dixon:szego:1978,picheny:wagner:ginsbourger:2012}.  Again~\cite{simulationlib} provides a convenient one-stop 
reference containing {\sf R} and {\tt MATLAB} code and visualizations.

\begin{align*}
 c_1(x) &= 0.5 \sin(2\pi(x_1^2 - 2x_2)) + x_1 + 2x_2 - 1.5\\
 c_2(x) &= -x_1^2 - x_2^2 + 1.5\\
  c_3(x) &= 15 - \left(15 x_2 -  \frac{5}{4 \pi^2} \left(15 x_1 - 5\right)^2 + \frac{5}{\pi} \left(15 x_1 - 5\right) - 6 \right)^2\!\! - 10 \left(1 - \frac{1}{8 \pi}\right) \cos(15 x_1 - 5)\\
 c_4(x) &= 4 - \left(4-2.1\left(2 x_1 - 1 \right)^2+\frac{\left(2 x_1 - 1 \right)^4}{3} \right)\left(2 x_1 - 1 \right)^2 - \left(2 x_1 - 1 \right)\left(2 x_2 - 1 \right) \\
 & - 16 \left(x_2^2 -x_2 \right) \left(2 x_2 - 1 \right)^2-  3\sin\left[12\left(1-x_1\right)\right] -  3\sin\left[12\left(1- x_2\right)\right]\\
 c_5(x) &= 3  +20 \exp\left(-0.2 \sqrt{\frac{1}{4} \sum_{i=1}^4(3x_i-1)^2}\right) + \exp \left(\frac{1}{4} \sum_{i=1}^4\cos(2\pi(3 x_i-1))\right) -\!20\!-\!\exp(1)\\
 c_6(x) &= \frac{1}{0.8387} \left[-1.1 + \sum_{i=1}^4{ C_i \exp \left( -\sum_{j=1}^4{  a_{ji} \left( x_j - p_{ji} \right)^2  } \right) } \right],
\end{align*}
with:
\begin{eqnarray*}
\mathbf{C} = \left[\!\begin{array}{c} 
1.0\\ 1.2\\ 3.0\\ 3.2 
\end{array} \!\right]\!\!, & 
\!\mathbf{a} = \left[\!\begin{array}{cccc}
           10.00 &  0.05&  3.00& 17.00 \\
           3.00& 10.00&  3.50&  8.00\\
           17.00& 17.00&  1.70&  0.05\\
           3.50&  0.10& 10.00& 10.00
\end{array}\!\right]\!\!, &
\!\mathbf{p} = \left[\! \begin{array}{cccc}
           0.131& 0.232& 0.234& 0.404\\
           0.169& 0.413& 0.145& 0.882\\
           0.556& 0.830& 0.352& 0.873\\
           0.012& 0.373& 0.288& 0.574
\end{array} \!\right]\!\!.
\end{eqnarray*}

%\pagebreak
\medskip
\noindent {\bf Problems:}
%\vspace{-0.5cm}
\begin{eqnarray}
%  \text{(LGB)} & \min_{x \in[0,1]^2}  f_1(x) & \text{s.t.} \quad c_1(x) \leq 0, \  c_3(x) \leq 0  \label{eq:LGB}\\
 \text{(LSQ)} &  \min_{x\in [0,1]^2} f_1(x) &  \text{s.t.} \quad c_1(x) \leq 0, \  c_2(x) \leq 0 \label{eq:gramacypb}\\
 \text{(GSBP)} & \min_{x \in[0,1]^2}  f_2(x) & \text{s.t.} \quad c_1(x) \leq 0, \   c_2(x) = 0, \ c_3(x) = 0  \label{eq:GGBP}\\
 \text{(LAH)} & \min_{x\in [0,1]^4} f_1(x) &  \text{s.t.} \quad c_5(x) \leq 0, \  c_6(x) = 0 \label{eq:LAH}
\end{eqnarray}

\begin{figure}[ht!]
\centering
\vspace{-0.25cm}
\includegraphics[width=.49\textwidth,trim=0 25 0 30]{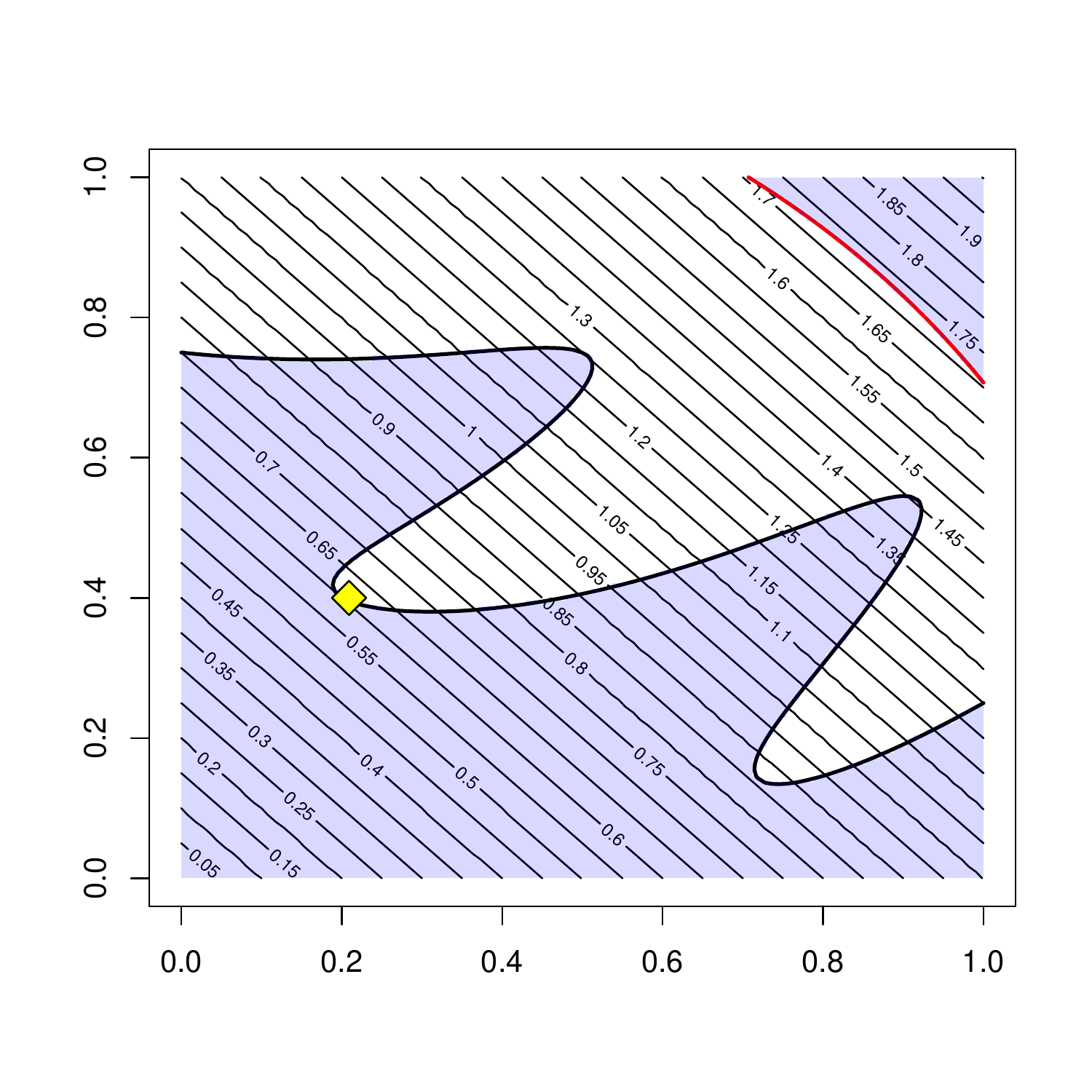}
\includegraphics[width=.49\textwidth,trim=0 25 0 30]{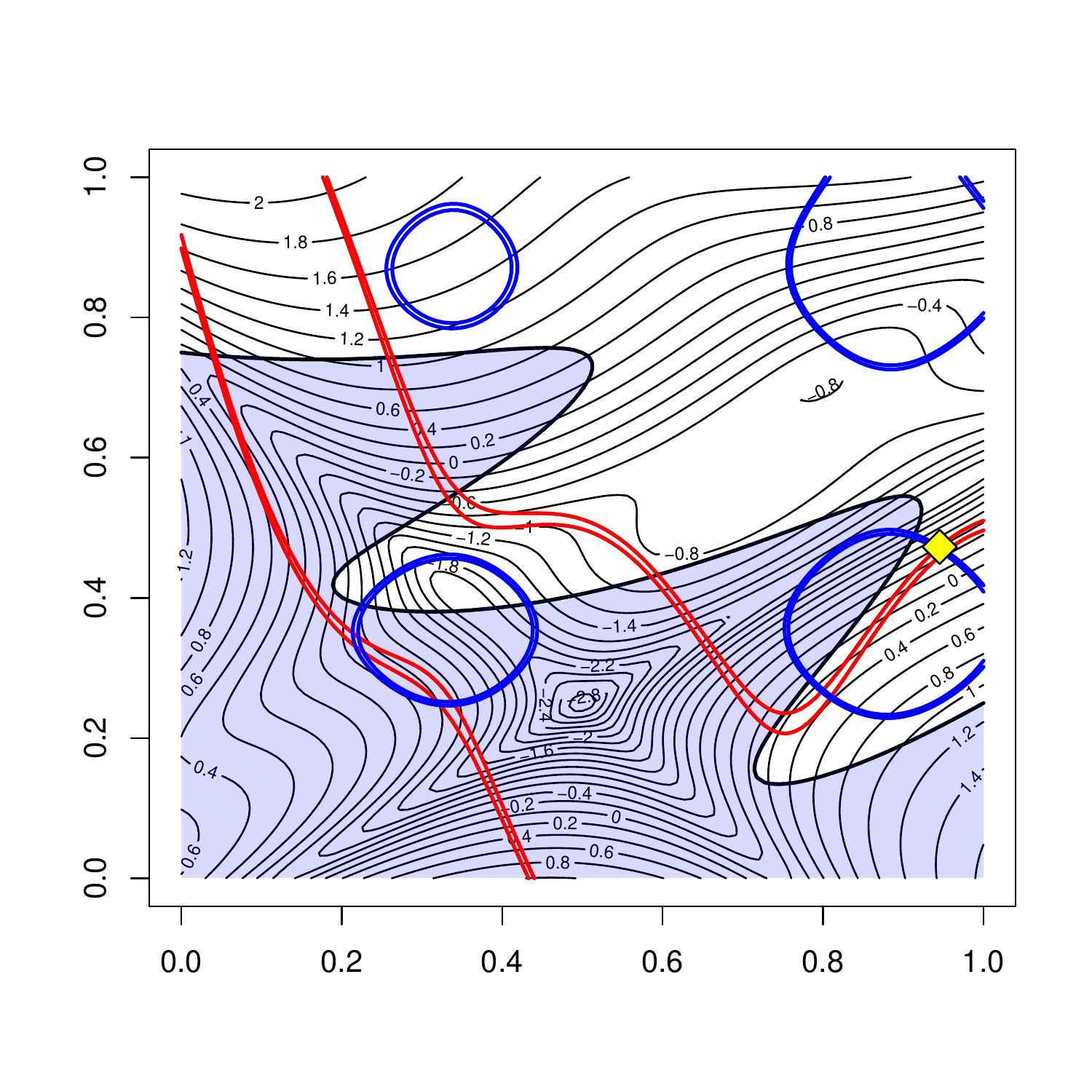}
\caption{LSQ (left) and GSBP (right) problems. The contour lines show the objective functions, 
 the bold lines the constraints, and the squares the solutions of the problems. Equality constraints are shown with two lines
 to represent the tolerance, the infeasible space defined by the inequality constraints are shaded.}
\label{f:testpbs}
\end{figure}

\noindent
Figure~\ref{f:testpbs} shows the two dimensional problems.

\bibliography{auglag}
\bibliographystyle{jasa}

\end{document}